\documentclass[journal]{vgtc}

\title{Mosaic Selections: Managing and Optimizing User Selections\\for Scalable Data Visualization Systems}

\author{
Jeffrey Heer, Dominik Moritz, Ron Pechuk
}

\authorfooter{
\item
  Jeffrey Heer is with University of Washington.
	E-mail: jheer@uw.edu.
\item
  Dominik Moritz is with Carnegie Mellon University.
	E-mail: domoritz@cmu.edu.
\item
  Ron Pechuk is with University of Washington.
	E-mail: rpechuk@uw.edu.
}

\shortauthortitle{: }

\abstract{Though powerful tools for analysis and communication, interactive visualizations often fail to support real-time interaction with large datasets with millions or more records.
To highlight and filter data, users indicate values or intervals of interest.
Such \emph{selections} may span multiple components, combine in complex ways, and require optimizations to ensure low-latency updates.
We describe Mosaic Selections, a model for representing, managing, and optimizing user selections, in which one or more filter predicates are added to queries that request data for visualizations and input widgets.
By analyzing both queries and selection predicates, Mosaic Selections enable \emph{automatic optimizations}, including pre-aggregating data to rapidly compute selection updates.
We contribute a formal description of our selection model and optimization methods, and their implementation in the open-source Mosaic architecture.
Benchmark results demonstrate orders-of-magnitude latency improvements for selection-based optimizations over unoptimized queries and existing optimizers for the Vega language.
The Mosaic Selection model provides infrastructure for flexible, interoperable filtering across multiple visualizations, alongside automatic optimizations to scale to millions and even billions of records.} 


\graphicspath{{figs/}{figures/}{pictures/}{images/}{./}} 

\usepackage{amsmath,amssymb}
\usepackage{microtype}                 
\PassOptionsToPackage{warn}{textcomp}  
\usepackage{textcomp}                  
\usepackage{mathptmx}                  

\usepackage{multirow}
\usepackage{colortbl}

\begin{document}



\maketitle

\section{Introduction}
\label{sec:intro}

To support effective analysis, interactive data systems should respond to user input at ``rates resonant with the pace of human thought'' \cite{doi:10.1145/2133806.2133821, doi:10.1109/TVCG.2014.2346452}, as ``milliseconds matter'' \cite{doi:10.1037/1076-898X.6.4.322} for user interface latency.
To ensure perception of continuity and causality, foundational texts in Human-Computer Interaction \cite{psych-hci} draw upon prior psychological studies to advocate interactive response rates under 100ms.
Experiments \cite{doi:10.1109/TVCG.2014.2346452, doi:10.1109/TVCG.2016.2607714} find that adding 500ms of latency to interactive visualizations can alter analysts' strategies and degrade performance, leading to reduced exploration and hypothesis generation.
However, not all interactive operations are equal, as the degree of latency sensitivity may vary \cite{doi:10.1111/cgf.13678, doi:10.1109/TVCG.2014.2346452}.
User performance may be robust to a one second delay when creating a new chart, whereas the same latency for navigation operations such as panning and zooming can render an application unusable.
In particular, updates for \emph{linked selections}---also known as \emph{brushing and linking} \cite{doi:10.1080/00401706.1987.10488204}---have been shown to be especially latency sensitive \cite{doi:10.1109/TVCG.2014.2346452}.

Though there are many expressive tools for visualization and data applications, most fail to
scale to large datasets while preserving low-latency interactions \cite{doi:10.1109/TVCG.2020.3028891, doi:10.1145/3639276}.
While researchers have developed methods for low-latency updates in response to dynamic user selections \cite{doi:10.1145/2882903.2882919, doi:10.1109/TVCG.2013.179, doi:10.1111/cgf.12129, doi:10.14778/3407790.3407826, doi:10.1145/3290605.3300924, doi:10.1111/cgf.13708, doi:10.1109/TVCG.2020.3030372},
most support only a limited subset of visualization types (such as scatter plots or time series) and interactions (such as panning and zooming).
Moreover, many optimizations are not designed to work together in a single system, and require manual tuning or precomputation.
Instead, we contend that interface designers and data analysts should be able to specify desired views and interactions at a high-level, and then benefit from automatic optimizations that, when possible, provide low-latency interaction ($\le$ 100ms) over backing data volumes with millions or even billions of records.

\begin{figure*}[t]
\centering
\includegraphics[width=\linewidth]{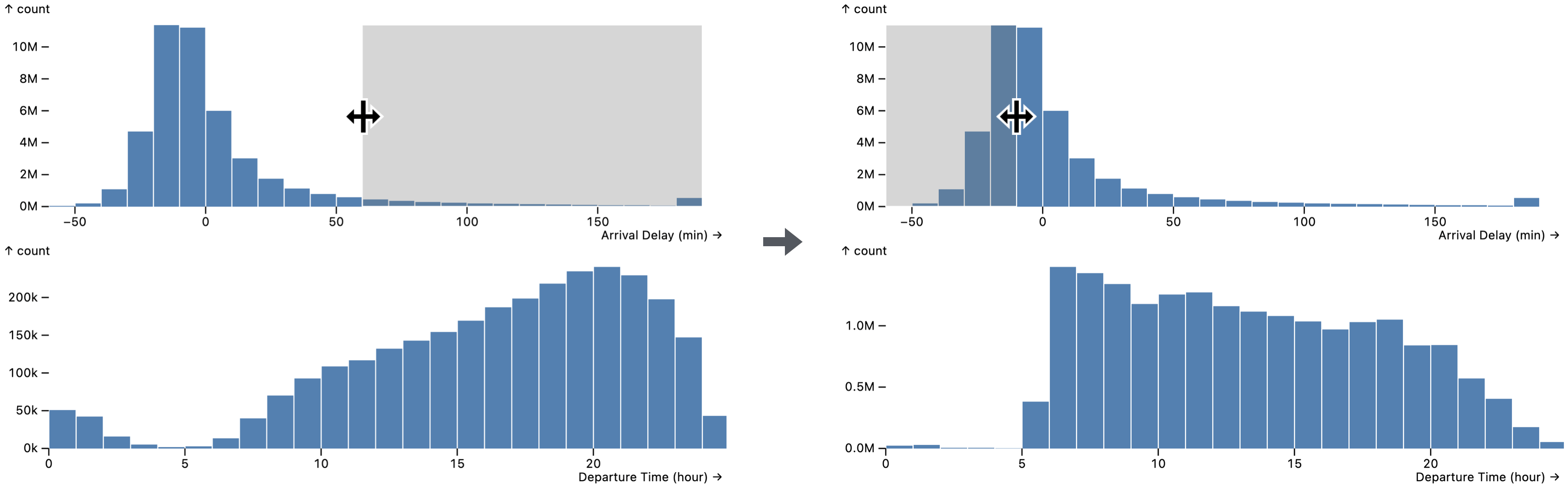}\vspace{-6pt}
\caption{Histograms of flight arrival data, interactively filtered by arrival delay intervals.}\vspace{-12pt}
\label{fig:example-flights}
\end{figure*}

We contribute a \textbf{formal model of user selections} in terms of \emph{client views} that consume data, \emph{interactors} that generate predicate clauses to select data subsets, and \emph{selections} that manage clause sets to resolve view-specific filters.
The selection model translates common operations performed in data analysis interfaces (using inputs such as menus, sliders, table views, and interactive visualizations) to filtering predicates applicable in database queries, and enables coordinated updates across multiple interface components.
Unlike many prior formulations \cite{hong2003compound, doi:10.1145/1357054.1357203, doi:10.1109/TVCG.2016.2599030}, our model flexibly supports \emph{cross-filtering} configurations in which selections may apply to some interface views, but not others.

As datasets grow to millions or more rows, rendering all records becomes infeasible due to both perceptual and computational constraints \cite{doi:10.1111/cgf.12129}.
Instead, it is common to bin, group, and aggregate data to achieve more scalable displays, ranging from basic histograms to high-resolution raster (pixel grid) views.
Interactive updates must bin, filter and re-aggregate data, leading to prohibitively high latency if performed against a full dataset upon every update.
To speed up these queries, \textbf{the selection model automatically optimizes updates using materialized views of pre-aggregated data}.\footnote{A materialized view stores the result of a database query by name.}
By analyzing client queries and a current active selection clause, the model determines appropriate dimensions and pre-aggregated measures to support efficient interactive updates.
Querying the materialized views can make queries multiple orders of magnitude faster, enabling low-latency updates.

General multi-query optimization requires queries ahead of time and is NP-hard \cite{doi:10.1145/3034786.3034792}.
In contrast, queries for interactive visualization applications are predictable \cite{doi:10.1111/cgf.12129, doi:10.1145/3290605.3300924}.
The selection model leverages an application's structure to make multi-query optimization and answering queries using views \cite{doi:10.1007/s007780100054} tractable.
We detail a materialized view construction and query approach, which unlike prior methods \cite{doi:10.1111/cgf.12129, doi:10.1145/3290605.3300924} supports aggregate operations beyond \texttt{COUNT} and \texttt{SUM}, and enables in-database computation and reuse across sessions and users.

We implemented the selection model and optimizations in Mosaic \cite{doi:10.1109/TVCG.2023.3327189}, an open-source architecture for linking databases and interactive views.
By default Mosaic leverages DuckDB \cite{doi:10.1145/3299869.3320212} for flexible deployment in web browsers, Jupyter notebooks, or using database servers.
An earlier paper \cite{doi:10.1109/TVCG.2023.3327189} presents the overall Mosaic architecture, describes included visualization and input widget components, and briefly introduces the selection model.
We extend that work with a formal selection model, implementation details, and pre-aggregation optimizations with expanded aggregate function support.
We also present performance benchmark results demonstrating significant improvements over naïve methods and existing systems, enabling a wider array of real-time interactions with large datasets.

In sum, this paper contributes
(1) a formal model of user selections that bridges user interactions and backing databases;
(2) optimizations that create and query pre-aggregated materialized views---applied \emph{automatically} based on query and selection clause analysis---to accelerate aggregate queries for rapid interactive updates; and
(3) performance benchmarks showing that Mosaic selections can reduce interactive latency by multiple orders of magnitude, outperforming existing systems.

\section{Motivation and Examples}
\label{sec:examples}

We first present motivating examples, illustrating interactive data systems ranging from basic visualizations to multi-view dashboards.
In each case, a graphical interface presents, filters, and highlights data queried from a backing relational database management system.
Interactive \emph{selections}---defined over discrete \emph{points} or continuous \emph{intervals}---indicate records of interest, which may then be filtered or highlighted within the interface.
Selection updates issue new queries to filter and potentially re-aggregate data.

\subsection{On-Time Flight Statistics}
\label{sec:on-time-flight-statistics}

\begin{figure}[t]
\centering
\vspace{4pt}{\small \begin{verbatim}
1  SELECT
2    24 * FLOOR(time / 24) AS x,
3    COUNT(*) AS y
4  FROM flights
5  WHERE delay BETWEEN $delay_i AND $delay_j
6  GROUP BY x
\end{verbatim}

}\vspace{-9pt}\vspace{-6pt}
\caption{Basic flight time histogram update query, filtered by an interval selection over delay values.}\vspace{-12pt}
\label{fig:flights-query}
\end{figure}

Figure~\ref{fig:example-flights} shows linked histograms for 44.8 million U.S. flights \cite{bts-ontime} from January 2018 to July 2024.
The top histogram shows flight delays, grouped into 10 minute bins.
The bottom histogram shows departure times in the local timezone, binned by hour.
Selecting an interval in one chart \emph{cross-filters} the other: the selection filters other plots, but not the plot in which the selection was made.
On the left, an interval selects flights delayed an hour or more; the time histogram reveals that these flights are likely to leave later, as delays compound throughout the day.
When a user moves the interval to select flights that arrived at least 10 minutes early (right), the departure times shift earlier in the day.

Figure~\ref{fig:flights-query} lists a basic SQL query for retrieving updated time histogram values in response to an arrival delay selection interval.
The query scans the full data to filter, bin, group, and count values.
Though this dataset fits in memory, such update queries can take over 100ms to complete, hampering analysis at the ``speed of thought.''
However, with appropriate optimizations---such as querying pre-aggregated materialized views---the latency drops to \textasciitilde{}1ms.
This paper contributes a model of user selections and query analysis to \emph{automatically} perform such optimizations and scale interactive data applications.

\subsection{New York City Taxi Trips}
\label{sec:new-york-city-taxi-trips}

\begin{figure}[t]
\centering
\includegraphics[width=\linewidth]{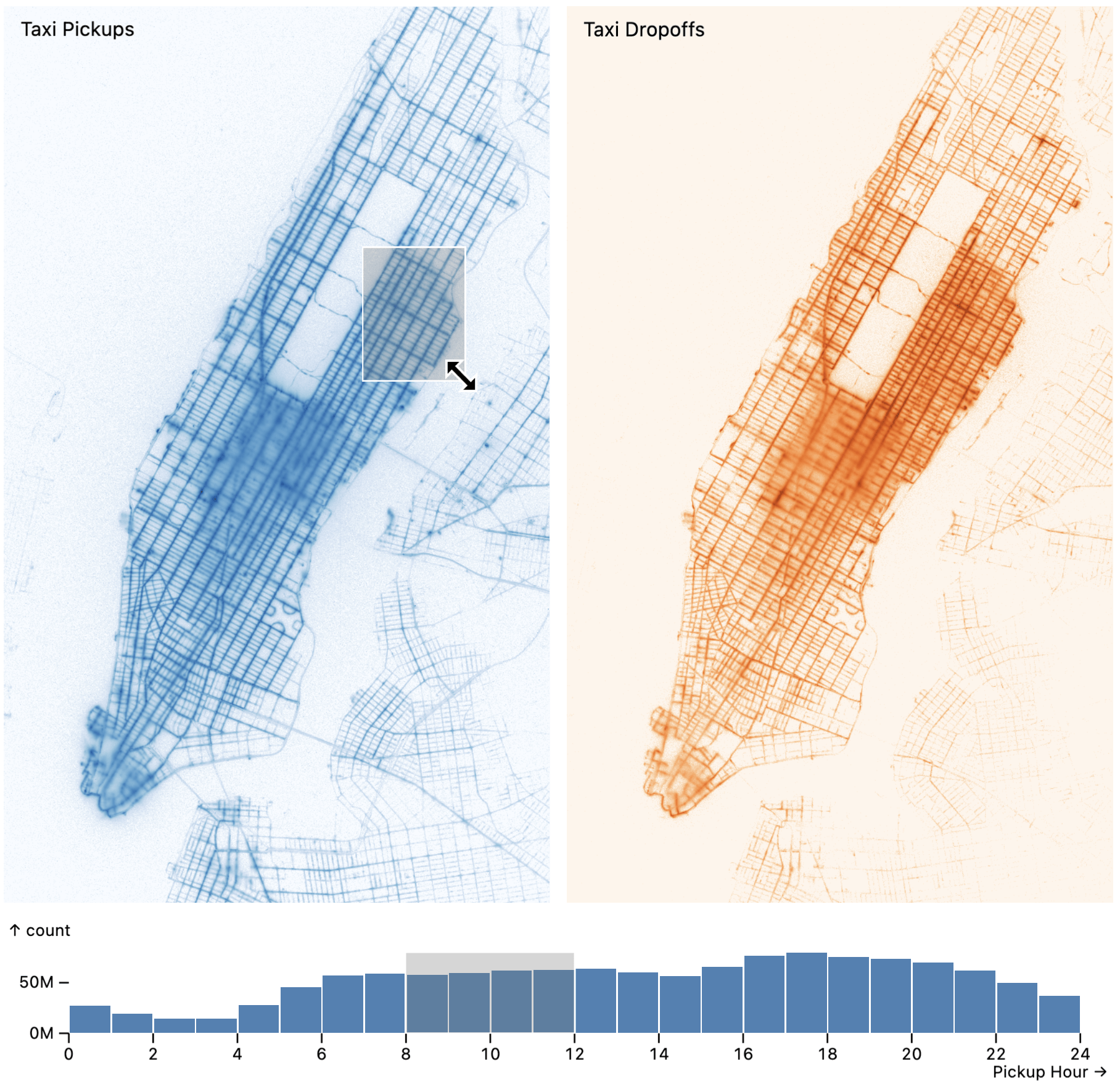}\vspace{-6pt}
\caption{Interactive maps of taxi pick-ups and drop-offs in New York City, cross-filtered by pickup time and location.}\vspace{-12pt}
\label{fig:example-taxi}
\end{figure}

Figure~\ref{fig:example-taxi} visualizes pickup and dropoff points for 1.3 billion New York City taxi trips from December 2008 to June 2016.
A histogram of pickup times additionally shows activity throughout the day.
All plots are interactive, enabling cross-filtering.
Here, a 1D interval selection in the time histogram isolates pickups during the hours of 8am--noon.
At the same time, a 2D interval selects taxi pickups in the Upper East Side, filtering the map of drop-offs on the right.
Perhaps surprisingly, we can see many taxi rides are local, with dropoffs primarily in the same region of town as pickups.

In this configuration, each visualization requires a different query.
With the current selections in Figure~\ref{fig:example-taxi}, the data for the pickups map is filtered by the pickup hour, while the data for the dropoffs map is filtered by both hour and pickup locations.
Evaluating both queries against the full data for each interactive update leads to high latency, requiring optimizations such as pre-aggregation.

\subsection{Protein Design Dashboard}
\label{sec:protein-design-dashboard}

\begin{figure}[t]
\centering
\includegraphics[width=\linewidth]{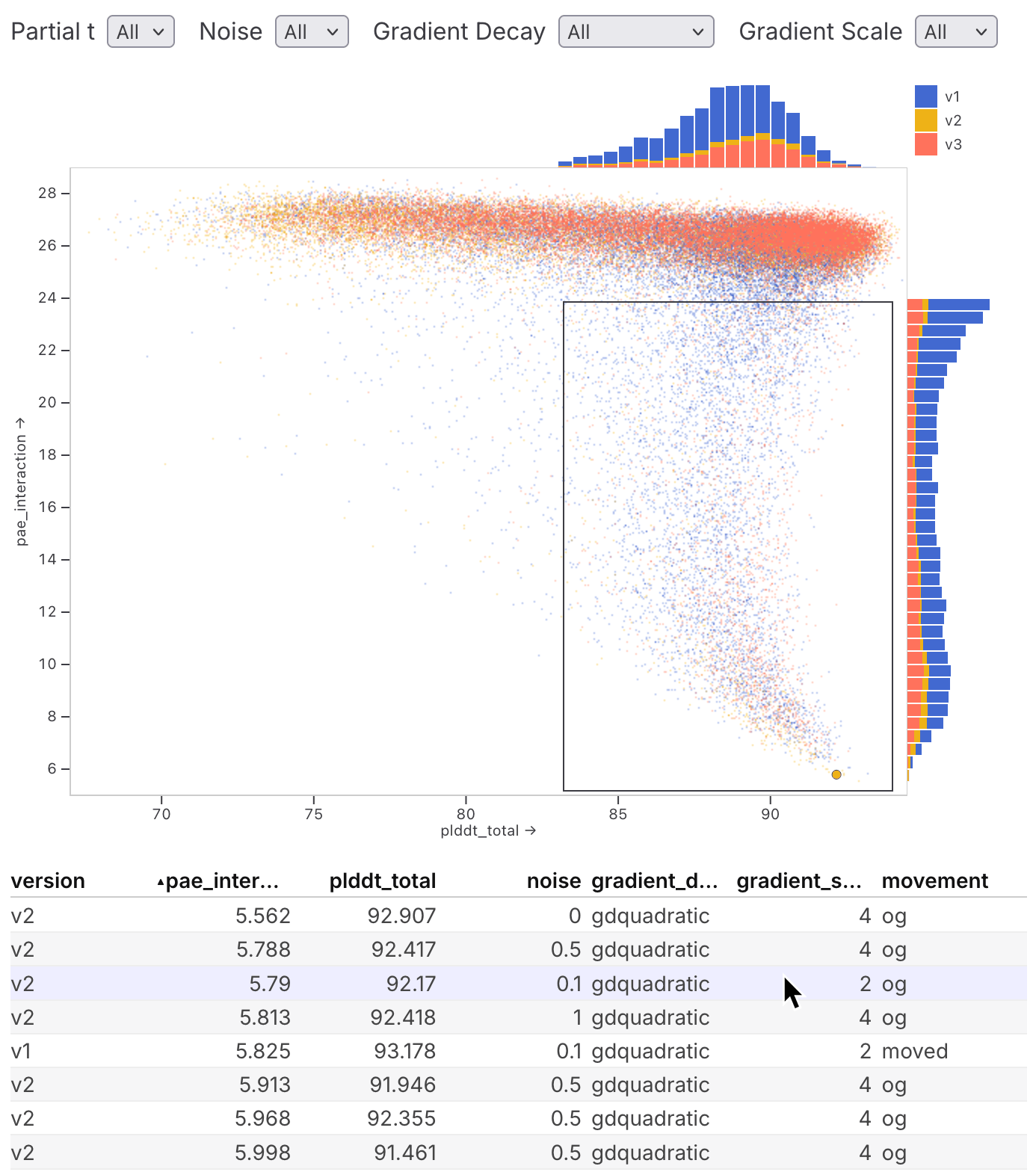}\vspace{-6pt}
\caption{Dashboard of synthetically designed proteins, with the most promising designs selected.}\vspace{-12pt}
\label{fig:example-proteins}
\end{figure}

Figure~\ref{fig:example-proteins} shows a dashboard of proteins synthesized using an AI diffusion model \cite{doi:10.1038/s41586-023-06415-8}, with process parameters and evaluation criteria for 48,000 proteins.
Drop-down menus enable filtering by process parameters.
An interactive legend enables isolation of results generated from a specific seed protein.
The central chart plots proteins by two evaluation scores, adorned with histograms of marginal distributions over the currently selected region.
A solid colored point highlights an individual protein, based on a hover selection in the table view below.

The dashboard involves many queries.
Unique values for process parameters are queried to populate the input menus.
The scatter plot and histograms each require their own queries, while the sortable table view queries for individual protein data.
This dashboard also involves a network of interactive selections.
The top input widgets filter the scatter plot and table views below.
In the scatter plot, a 2D interval selection filters data for both the table and histograms.
Finally, selecting rows in the table highlights scatter plot points for individual proteins.
Some standard optimizations can be applied here; for example, the table component limits queries to data around the visible scroll window.
As the scatter plot has a fixed domain (the axes do not change across selections), it is also possible to perform pre-aggregation of sufficient statistics for the histograms---binned at the pixel level---to optimize interactive filtering as the 2D selection interval changes (details in \S\ref{sec:opt}).

\subsection{Gaia Star Catalog}
\label{sec:gaia-star-catalog}

\begin{figure}[t]
\centering
\includegraphics[width=\linewidth]{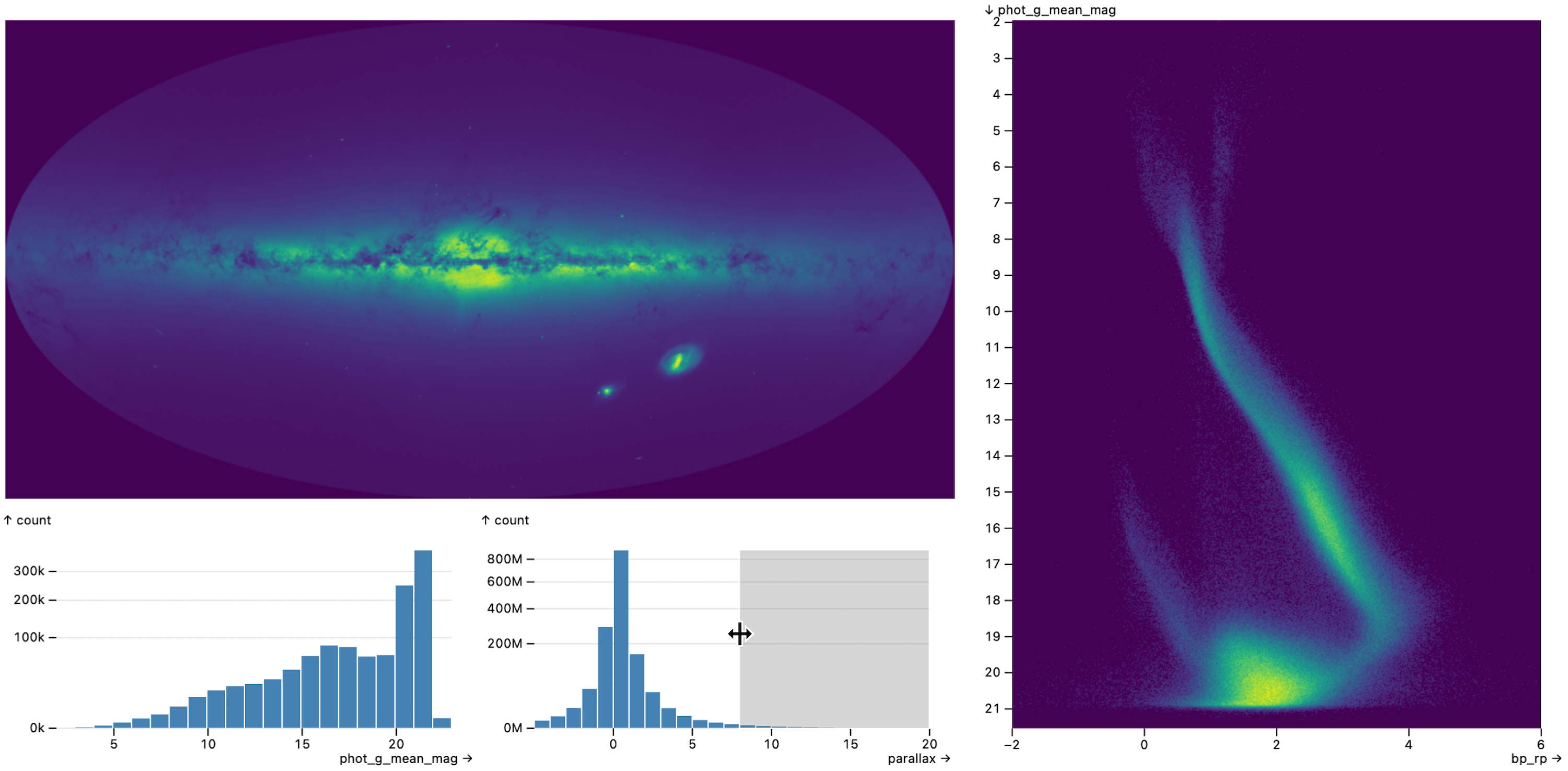}\vspace{-6pt}
\caption{Linked visualizations of over 1.8 billion stars in the Gaia star catalog, with high parallax stars selected.}\vspace{-12pt}
\label{fig:example-gaia}
\end{figure}

The Gaia catalog is a survey of over 1.8B stars \cite{gaia:gdr3}.
Figure~\ref{fig:example-gaia} contains a high-resolution density map of projected sky coordinates, showing the Milky Way and satellite galaxies.
Histograms of magnitude and parallax, alongside a density plot of star color vs. stellar magnitude (a \href{https://en.wikipedia.org/wiki/Hertzsprung-Russell_diagram}{Hertzsprung-Russell diagram}), show additional data.
Interactive selections in any one view cross-filter the others.
Here, a 1D interval selection selects stars with higher parallax values.
Upon filtering, the raster plot of color versus stellar magnitude reveals a curve tracing from the upper-left to the bottom-right: a common trajectory of stellar evolution known as the ``main sequence.''
Meanwhile, the large mass of stars towards the bottom suffers from data quality issues \cite{gaia-release}.

In this configuration, any one view is filtered by the intersection of selections from the others.
Updates for each visualization can be optimized using materialized views of pre-aggregated data.
The most demanding pair of views is the sky map and Hertzsprung-Russell diagram: both involve high-resolution raster displays, where visual encodings and selections operate over pixel-level bins.
If we assume each view has $300 \times 200$ pixels (bins), a dense representation of per-bin pre-aggregated data requires $(300 \times 200)^2$ = 3.6B values.
Techniques that materialize a dense pre-aggregated view, such as imMens \cite{doi:10.1111/cgf.12129} and Falcon \cite{doi:10.1145/3290605.3300924}, run out of memory and fail at this scale.
Fortunately, due to sparsity (in this case over four dimensions) many fewer values are actually needed; the automatic optimization methods presented in this paper use a sparse representation that accounts for this.

\section{Related Work}
\label{sec:related-work}

The Mosaic selection model draws on prior work in visualization systems, interactive selections, and database optimization.

\subsection{Visualization Systems}
\label{sec:visualization-systems}

Popular visualization frameworks support an array of visual encodings.
For example, Wilkinson's Grammar of Graphics \cite{doi:10.1007/978-3-642-21551-3_13} helped popularize a compositional approach to visual encoding rather than simple taxonomies of chart types, inspiring widely used tools such as ggplot2 \cite{doi:10.1198/jcgs.2009.07098}.
Web-based tools such as D3 \cite{doi:10.1109/TVCG.2011.185}, Vega \cite{doi:10.1109/TVCG.2015.2467091}, and Vega-Lite \cite{doi:10.1109/TVCG.2016.2599030} further provide rich support for \emph{interactive} visualization, through operations such as panning, zooming, and linked selections for filtering and highlighting.
Tableau (previously Polaris \cite{doi:10.1109/2945.981851}) is a commercial tool that uses a Grammar of Graphics approach and also queries databases to scale to larger datasets.
However, these tools fail to provide low-latency interactions as dataset sizes grow to millions or more rows.

VegaFusion \cite{doi:10.1109/VIS54862.2022.00011} and VegaPlus \cite{doi:10.1145/3639276} analyze Vega \cite{doi:10.1109/TVCG.2015.2467091} workflow specifications and improve scalability by rewriting a Vega dataflow to query a relational database.
They do not, however, provide pre-aggregation optimizations.
As shown in Figure~\ref{fig:benchmark-results}, even unoptimized Mosaic selection update queries outperform these systems.

Some visualization systems scale to larger dataset sizes by focusing on specific visualization and/or interaction types.
ForeCache \cite{doi:10.1145/2882903.2882919} leverages a machine learning model trained on user navigation behavior to prefetch data subsets for multi-scale pan/zoom navigation.
Kyrix \cite{doi:10.1111/cgf.13708} targets scalable zoomable user interfaces, including precomputation of spatial positions and indexes.
Kyrix-S \cite{doi:10.1109/TVCG.2020.3030372} extends this work with a focus on scatter plot visualizations.

Other methods optimize filtering and aggregation.
Nanocubes \cite{doi:10.1109/TVCG.2013.179} are spatio-temporal indexes for low-latency selection updates, but can suffer from long construction times (3--4 hours for 1B rows).
imMens \cite{doi:10.1111/cgf.12129} uses precomputed, pre-aggregated data tiles, represented as dense n-dimensional arrays, to provide rapid client-side updates for \texttt{COUNT} and \texttt{SUM} aggregates.
Falcon \cite{doi:10.1145/3290605.3300924} builds on this approach, supporting higher-resolution selections via pixel-level binning, prefetching of data tiles based on cursor position, and summed area tables for constant time update queries.
Khameleon \cite{doi:10.14778/3407790.3407826} streams approximate data tiles with a scheduler that trades off result quality and bandwidth; such techniques might be adapted to the context of Mosaic selections.

Mosaic selections generalize and improve upon imMens and Falcon.
Instead of dense arrays, Mosaic uses pre-aggregated materialized views realized as sparse database tables.
Mosaic also supports many more aggregation functions (Table~\ref{tbl:supported-aggs}).
While imMens and Falcon require developers to explicitly orchestrate optimizations, Mosaic Selections instead apply pre-aggregation optimizations \emph{automatically}.

In contrast to prior work, we contribute a general model of user selection applicable to varied visualization and interactor implementations.
The Mosaic selection model enables \emph{automatic} optimization techniques for low-latency interaction over aggregated data by analyzing both client view queries and the active selection clause.
On commodity laptops with datasets ranging up to 100M records (Figure~\ref{fig:benchmark-results}), these optimizations can be applied immediately without a prior precomputation phase, enabling ``cold start'' visualization and exploration.

\subsection{Modeling User Selections}
\label{sec:modeling-user-selections}

Prior work has modeled user selections as predicate functions, typically in the context of a visualization environment.
Snap-Together visualization \cite{doi:10.1145/345513.345282} provides a limited yet easy-to-use model of view coordination based on primary key relations.
DEVise \cite{doi:10.1145/253262.253335} and later Improvise \cite{weaver2004improvise} map user selections such as points or intervals to query predicates for highlighting and filtering.
VQE (Visual Query Environment) \cite{doi:10.1145/263407.263545} and generalized selection via query relaxation \cite{doi:10.1145/1357054.1357203} follow a similar model.
VIQING (Visual Interactive QueryING) \cite{olston1998viqing} introduces a notion of visual joins, but does not support multi-view coordination.
Compound brushing \cite{hong2003compound} provides a visual programming tool to define custom selection behavior.
DIEL \cite{doi:10.1109/TVCG.2021.3114796} models interactions as asynchronous event streams that are then captured and queried in event logs, but requires writing low-level application-specific queries and library integrations.

Perhaps most similar to the Mosaic selection model is the \emph{selection} abstraction of Vega-Lite \cite{doi:10.1109/TVCG.2016.2599030}, which couples both event handling and predicate generation for a single input source.
The Mosaic selection model instead decouples input event handling, to manage and resolve predicate clauses from diverse and extensible input sources (\emph{interactors}), ranging from basic widgets to visualizations.
Unlike Vega-Lite and other prior works, the Mosaic model handles more complex coordination behaviors such as cross-filtering, while providing automatic optimization for aggregation queries.

\subsection{Relevant Database Optimizations}
\label{sec:relevant-database-optimizations}

The use of pre-aggregated materialized views is inspired by the database \texttt{CUBE} operator \cite{datacube}.
Harinarayan et al.~\cite{doi:10.1145/235968.233333} describe how to implement data cubes efficiently.
They introduce a lattice framework to model complex groupings that involve arbitrary hierarchies of attributes, and then propose a near optimal algorithm to select materialized views.
Multiple problems arise when using pre-aggregation: which views to materialize and how to query the materialized views.
In our context, we do not need to select from a full lattice, and can leverage the selection model to automatically determine which materialized views are needed.
Rather than build a full cube, Mosaic materializes cube \emph{projections} specific to a (selection clause, client view) pair.
Relative to the general problems of answering queries using views \cite{doi:10.1007/s007780100054} and multi-query optimization \cite{doi:10.1145/342009.335419, doi:10.1145/42201.42203}, our selection model simplifies the choice of which materialized view to query.
While DuckDB does not yet provide internal support for materialized views, in other contexts our model could leverage incremental view maintenance \cite{doi:10.7551/mitpress/4472.003.0016} to keep pre-aggregations up to date after changes to a base table.

\section{A Model of User Selections}
\label{sec:sel}

We assume an interactive data system consists of a set of \emph{client views}\footnote{Here we use the term ``view'' in the sense of a visible interface component, not in the database sense of a logical relation defined by a query.} ($v \in V$) that consume data queried from a relational database.
A view $v$ is a data consumer, which represents its data needs as a declarative query $q_v$.
Given a boolean-valued predicate $p$, the view requests needed data by providing a filtered query $q_{vp}$ that incorporates $p$.
In Figure~\ref{fig:flights-query}, the addition of line 5 produces $q_{vp}$ from an initial $q_v$.
Query results are later returned to $v$, to process and display the results as desired.

Client views in a data interface can include formatted text (e.g., to present a summary statistic), input widgets (such as menus, sliders, or text entry), or data displays such as tables and visualizations.
For example, a menu widget may generate a query $q_v$ for all distinct values of a column in a table (filtered by an input predicate $p$), to include within the menu.
Similarly, a text search widget might query for distinct values to populate auto-complete suggestions.
An interface component may consist of one or more client views.
Visualizations often have multiple layers, each corresponding to a view $v_i$ with different data needs.

\begin{figure}
\centering
\includegraphics[width=\linewidth]{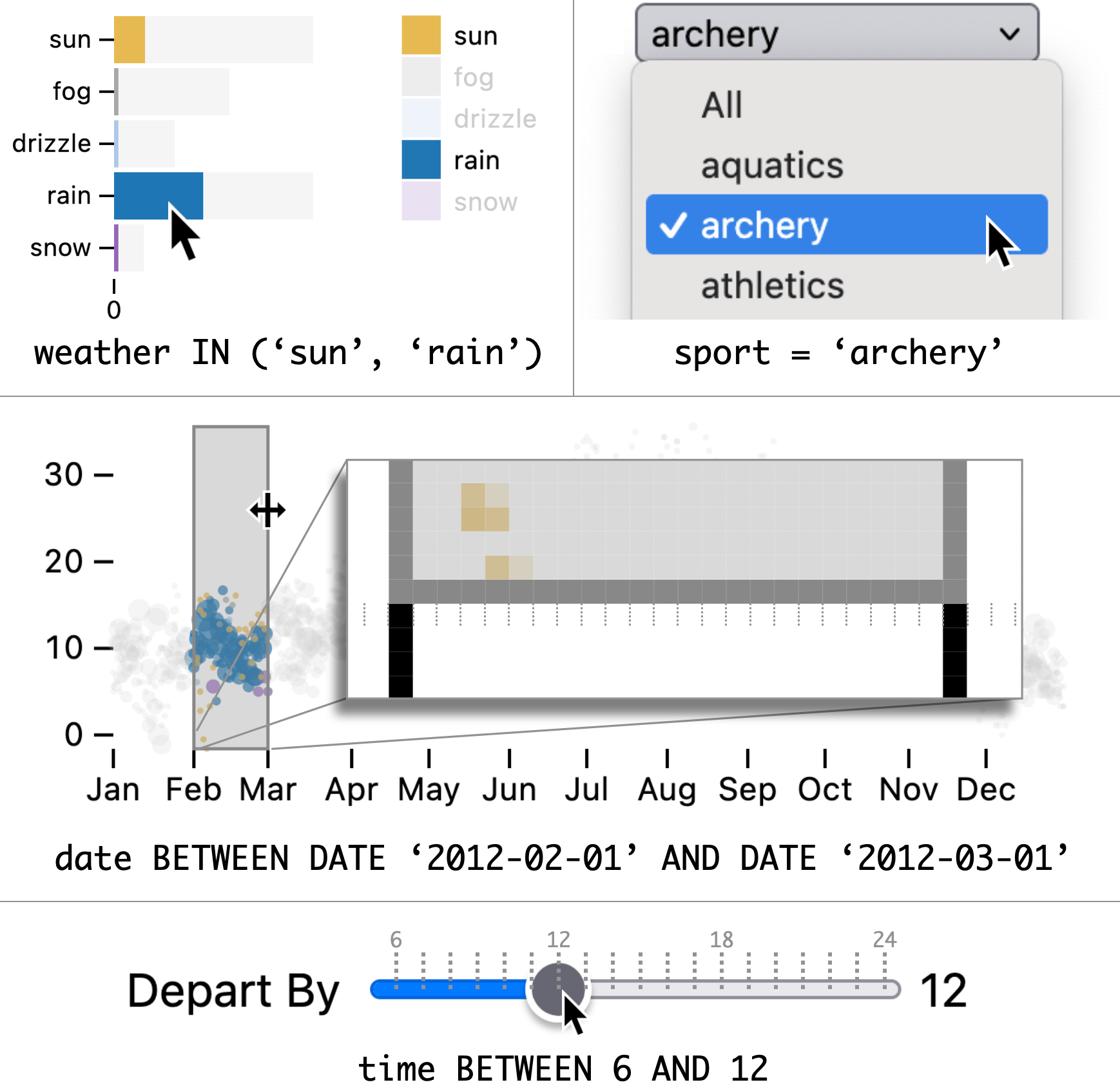}
\vspace{-10pt}\vspace{-6pt}
\caption{Interactors with predicate clauses. Top: \emph{point} selections based on chart elements, legend entries, or menu items. Middle: 1D \emph{interval} selection over a chart axis. Bottom: slider-based \emph{interval} selection. Dotted lines indicate possible discrete end points for interval selections.}\vspace{-6pt}
\label{fig:interactors}
\end{figure}

An \emph{interactor} is a predicate clause generator: a component that generates boolean-valued predicate clauses used to filter query results for one or more views (Figure~\ref{fig:interactors}).
For example, selecting items in a list widget triggers an interactor to generate a clause that matches just the selected items.
Similarly, as a user drags along a chart axis to indicate an interval, an interactor generates a corresponding range predicate clause over the selected interval.
Views and interactors are often tightly coupled, but this is not a strict requirement.
A view for a scatter plot visualization might use multiple interactors, each generating distinct selection clauses (e.g., one for mouse hover selections and another for click-drag interval selections).
Meanwhile, a text display view that simply prints a summary statistic may have no interactor at all.

\subsection{Selections: Clause Sets and Resolution}
\label{sec:sel-def}

Formally, a selection $S = \langle C, R \rangle$ consists of a set of predicate clauses $C$ provided by interactors and a resolution operator $R: \langle C, v \rangle \rightarrow p$.
Each clause $c \in C$ corresponds to a boolean-valued selection predicate, as one might find in a SQL \texttt{WHERE} clause.
For example, if a user selects the interval $[x_i, x_j] \in X$ in one visualization and the category value $y_k \in Y$ in another, the resulting clause set might be:

$C = \lbrace (X \texttt{\;BETWEEN\;} x_i \texttt{\;AND\;} x_j), (Y = y_k) \rbrace$

The resolution operator $R$ takes a clause set $C$ and a view $v$ as arguments and produces an output predicate $p$ that can subsequently be applied to filter data for $v$.
The operator $R$ does so by aggregating boolean predicate clauses.
Basic resolution operators are \texttt{LAST} (include only the most recently added, or \emph{active}, clause), \texttt{INTERSECT} ($\cap$, boolean and) and \texttt{UNION} ($\cup$, boolean or).
For example, the use of \texttt{LAST} ensures that only the most recently added selection clause is applied, clearing the selection of any previous clauses.

These basic operators produce the same result regardless of the client view argument $v$.
As such, they are insufficient to support \emph{cross-filtering}, in which a selection clause filters other views, but not those associated with the originating interactor.
To support cross-filtering, the resolution operator must aggregate all clauses \emph{except} those for which the input view $v$ is associated with an initiating interactor.
Accordingly, a cross-filtering resolution operator must filter out related clauses in addition to aggregating clauses (e.g., via \texttt{INTERSECT} or \texttt{LAST}).

To appropriately update and cross-filter, a clause $c$ must include additional properties: a unique $source$ id of the originating interactor and a set of $views$ associated with the clause that should not be filtered in the case of cross-filtered clause resolution.
During cross-filtering, a clause whose $views$ set includes the input view $v$ will be omitted from the resolved predicate.
It is the responsibility of an interactor to provide this information as part of a generated clause.

\begin{table}[t]
\centering

\begin{tabular}{ll}
Name & Description \\
\hline
$predicate$ & A boolean-valued filter predicate. \\
$source$ & A unique id for the originating interactor. \\
$views$ & A set of views to exclude during cross-filtering. \\
$meta$ & Metadata key-value pairs to aid optimization (\S\ref{sec:opt}).
\end{tabular}

\vspace{-6pt}
\caption{Selection clause properties.}\vspace{-4pt}
\label{tbl:clause-props}
\end{table}

\begin{table}[t]
\centering

\begin{tabular}{ll}
Name & Description \\
\hline
$type$ & The selection type: \emph{interval}, \emph{point}, or \emph{match}. \\
$pixelSize$ & The interactive pixel resolution (\emph{interval} type only). \\
$bin$ & A bin function (default \texttt{FLOOR}, \emph{interval} type only). \\
$scales$ & An array of scale descriptors (see Table~\ref{tbl:scales}).
\end{tabular}

\vspace{-6pt}
\caption{Selection clause metadata properties.}\vspace{-12pt}
\label{tbl:clause-meta}
\end{table}

\begin{table}[t]
\centering

\begin{tabular}{ll}
Name & Description \\
\hline
$type$ & The scale type (\emph{linear}, \emph{log}, \emph{pow}, \emph{symlog}). \\
$domain$ & Start and end values $[d_0, d_1]$ in data space. \\
$range$ & Start and end values $[r_0, r_1]$ in screen space. \\
$base$ & The base of the logarithm (\emph{log} scales only). \\
$exponent$ & The power exponent (for \emph{pow} scales). \\
$symlog$ & The symmetric log constant (for \emph{symlog} scales) \cite{doi:10.1088/0957-0233/24/2/027001}.
\end{tabular}

\vspace{-6pt}
\caption{Scale descriptor properties, used in \emph{interval} clauses.}\vspace{-12pt}
\label{tbl:scales}
\end{table}

\subsection{Selection Clause Metadata}
\label{sec:sel-meta}

Table~\ref{tbl:clause-props} summarizes the properties of a selection clause.
In addition to the filtering \emph{predicate}, \emph{source} interactor, and client \emph{views} set, a clause may include metadata (\emph{meta}) to inform query optimization.
To optimize selection updates, we require an appropriate binning scheme that maps data values to discrete dimensions in the pre-aggregated materialized views.
To perform such binning in-database, sufficient information about data encodings must be passed from the interface.
The metadata properties document the selection's semantics and any associated data-space-to-screen-space transformations, such as the mapping between column values and screen pixel positions.
The interactor that generates the clause is responsible for providing these metadata values.

Table~\ref{tbl:clause-meta} lists selection clause metadata properties.
The \emph{type} property indicates if the clause operates over \emph{point} values (corresponding to equality comparisons), \emph{interval} values (corresponding to range queries), or \emph{match} criteria (for pattern matching, including text \emph{prefix}, \emph{suffix}, \emph{contains}, or \emph{regexp} searches).
A \emph{point} type indicates that the clause will take the form of a disjunction (logical or) of value equality checks.
In this case, the type information alone is sufficient for pre-aggregation, as each discrete point value corresponds to a bin.

Optimizing updates for an \emph{interval} selection requires additional metadata.
The \emph{pixelSize} metadata property indicates the size of an ``interactive'' pixel.
This value is configurable by interactors to trade-off interactive resolution with performance \cite{doi:10.1145/3290605.3300924}.
The \emph{pixelSize} defaults to 1, for a 1:1 mapping between screen pixels and pre-aggregate bins (as in Figure~\ref{fig:interactors}, middle).
Setting a larger value applies a grid with bins larger than screen pixels.
If the \emph{pixelSize} is 2, adjusting an interval endpoint within a 2-screen-pixel bin will have no effect (reduced resolution) but the materialized view may be half the size (hence faster to query).

The \emph{bin} property indicates how values should be discretized.
By default the \texttt{FLOOR} function is used.
\texttt{CEIL} and \texttt{ROUND} are also applicable, if indicated by an interactor.
For example, if using a slider input to select a thresholded $[min, value]$ interval (as in Figure~\ref{fig:interactors}, bottom), we use \texttt{CEIL} to ensure the interval is inclusive of the selected slider $value$.

Determining appropriate bins for an \emph{interval} selection requires knowledge of the scale transforms used to map data values to screen space values.
Table~\ref{tbl:scales} describes the requisite scale properties.
With a linear scale, binning is straightforward.
Given a value $x$ to bin, scale \emph{domain} $d$ (data space extent), and \emph{range} $r$ (screen space extent):

\vspace{-8pt}\begin{align}
\label{eqn:bin-linear}
bin(x) = \texttt{FLOOR} \left( \frac{r_1 - r_0}{pixelSize} \times \frac{x - d_0}{d_1 - d_0} \right)
\end{align}

Other common scale types, such as \emph{log}, \emph{symlog} (symmetric log, an alternative that supports negative and zero values \cite{doi:10.1088/0957-0233/24/2/027001}), and \emph{pow} (exponentiation), involve non-linear transformations.
These transformations are parameterized, either by the \emph{base} of the logarithm, the symlog \emph{constant}, or the power scale \emph{exponent}.
For example, to bin values plotted on a logarithmic scale, we use the formula:

\vspace{-8pt}\begin{align}
\label{eqn:bin-log}
bin(x) = \texttt{FLOOR} \left( \frac{r_1 - r_0}{pixelSize} \times \frac{log_{base}(x) - log_{base}(d_0)}{log_{base}(d_1) - log_{base}(d_0)} \right)
\end{align}

For a 1D \emph{interval} selection, only a single scale descriptor is needed.
For higher dimensional \emph{interval} selections (e.g., $n$=2), $n$ scale descriptors are needed, one for each dimension.
In sum, selection clauses include not only filtering predicates and cross-filtering information, but also encoding metadata for later optimization.

\section{Automatic Selection Optimization}
\label{sec:opt}

The selection model enables \emph{automatic} optimization of aggregate queries using pre-aggregated materialized views \cite{doi:10.1007/s007780100054}, similar to a data cube \cite{datacube}.
For each view $v$ with a compatible query $q_v$, we can construct a materialized view that supports update queries when an interactor repeatedly updates the active clause $c_a$ within a selection that uses \texttt{INTERSECT} or \texttt{LAST} to resolve clauses.
Formally, given a selection $S$ with active clause $c_a$ and a set of views $v \in V_S$ filtered by $S$, we seek to create and query tables with pre-aggregated data for all compatible view queries $q_v$.
The result is a unique materialized view for each (active clause, view query) pair $\langle c_a, q_v \rangle$.

Consider the flight data in Figure~\ref{fig:example-flights}.
When a user selects an interval in the top (delay) histogram, a predicate for the selected interval is provided by the active clause $c_a$.
As the user adjusts the interval extent $[x_i, x_j]$, the interval endpoints of each successive $c_a$ change, yet the predicate \emph{structure} is constant: $X \texttt{\;BETWEEN\;} x_i \texttt{\;AND\;} x_j$.
We leverage this structure to create a materialized view that can service all such updates.

To provide updates for the second (time) histogram, we construct a materialized view with three columns.
The first two columns are groupby \emph{dimensions}: the histogram bins (hour of day $h$) and bins for each possible interval endpoint of $c_a$, which corresponds to the pixel width of the chart plotting area (assuming \emph{pixelSize} = 1).
The third column is a \emph{measure}: the count of records in each $\langle c_a, h \rangle$ bin.
Figure~\ref{fig:pre-agg-sql} lists the SQL queries for creating and querying this table.
Line 5 of the creation query follows Equation~\ref{eqn:bin-linear}: the variable \texttt{\$pixels} is the pixel width of the x-axis ($r_1 - r_0$, $pixelSize$=1), while \texttt{\$min} and \texttt{\$max} comprise the x-axis scale domain $[d_0, d_1]$.

\begin{figure}[b]
\centering
\vspace{-8pt}{\small \begin{verbatim}
1  CREATE TABLE IF NOT EXISTS mosaic.pre_agg_a097caa4 AS
2  SELECT
3    24 * FLOOR(time / 24) AS x,
4    COUNT(*) AS y,
5    FLOOR($pixels * (delay - $min) / ($max - $min)) AS active
6  FROM flights
7  GROUP BY x, active

1  SELECT x, SUM(y) FROM mosaic.pre_agg_a097caa4
2  WHERE active BETWEEN $pixel_i AND $pixel_j
3  GROUP BY x
\end{verbatim}

}\vspace{-8pt}\vspace{-6pt}
\caption{Creation and update queries for a pre-aggregated materialized view over flight times, filtered by an interval selection over delay values.}
\label{fig:pre-agg-sql}
\end{figure}

The size of the materialized view is at most 14,400 rows (600 pixels $\times$ 24 hour bins)---over three orders of magnitude less than the original 44.8 million rows.
Moreover, some pixel bins are empty due to sparsity, so the actual number of rows in the materialized view is only 5,996.
As the interval selection changes, we can re-query the materialized view using the current pixel bin values of $c_a$ for rapid updates.

More generally, to automatically pre-aggregate data, we

\begin{enumerate}
\item check if a query $q_v$ filtered by active clause $c_a$ performs aggregation amenable to optimization (\S\ref{sec:opt-compat}),
\item determine dimension columns by extracting any groupby values referenced in $q_v$, as well as dimensions for the possible values of the active clause $c_a$ (\S\ref{sec:opt-groupby}), and
\item create a materialized view with measure columns containing sufficient statistics for all aggregate query results (\S\ref{sec:opt-stats}).
\end{enumerate}

\subsection{Materialized View Compatibility}
\label{sec:opt-compat}

To construct a materialized view, both the active clause $c_a$ and the view query $q_v$ must be amenable to analysis and compatible with pre-aggregation.
If any of the following checks fail, a materialized view is not constructed, and instead the query $q_{vp}$ is issued directly.

First, the selection $S$ must use \texttt{INTERSECT} or \texttt{LAST} aggregation in the resolution operator $R$.
Though \emph{point}-type clauses may internally include disjunctions, \texttt{UNION} resolution is not amenable to pre-aggregation.
The active clause $c_a$ must be either a \emph{point} or \emph{interval}-type clause, as \emph{match} clauses lead to an intractably large space of dimensions (all possible subset matches).
The active clause's \emph{meta} property must be defined as described in \S\ref{sec:sel-meta}.
For example, an \emph{interval} clause must indicate its \emph{type} and include \emph{scales} for determining pixel-level selection bins.

Next, the groupby dimensions in $q_v$ must form a fixed domain: the dimensions (such as data binning schemes) should not change as a result of applying a selection filter.
Consider a histogram that bins data values.
If the binning grid is fixed, the bins serve as dimensions for querying filter updates.
However, if applying a selection changes the binning scheme---e.g., by zooming in to a sub-region and re-binning to a finer resolution---the materialized view dimensions are unstable and hence unusable.
We can not determine if dimensions are stable across updates solely by examining $c_a$ and $q_v$ alone.
Instead the client view $v$ is responsible for providing this information: $v$ must include a \texttt{filterStable} property set to \texttt{true} if the groupby dimensions of $q_v$ are stable across provided filtering predicates $p$.

Finally, the view query $q_v$ must use supported aggregation operations over a single source relation.
View queries may include subqueries or common table expression (CTEs); examples include queries for hexagonal binning, linear binning for improved density estimation \cite{doi:10.1109/VIS49827.2021.9623323, doi:10.1109/TVCG.2023.3327189, doi:10.1080/00949658308810650}, and density line charts \cite{doi:10.48550/arXiv.1808.06019}.
We check $q_v$ by walking an abstract syntax tree of the query to ensure that only one named relation (table or view) serves as a source of aggregated data.

\begin{table}[b]
\centering
\vspace{-12pt}
\begin{tabular}{ll}
Category & Functions \\
\hline
Summary & \texttt{COUNT}, \texttt{SUM}, \texttt{MAX}, \texttt{MIN}, \texttt{PRODUCT} \\
Logical & \texttt{BIT\_AND}, \texttt{BIT\_OR}, \texttt{BIT\_XOR}, \texttt{BOOL\_AND}, \texttt{BOOL\_OR} \\
1D Stats & \texttt{AVG}, \texttt{GEOMEAN}, \texttt{ARG\_MAX}, \texttt{ARG\_MIN}, \\
 & \texttt{VAR\_SAMP}, \texttt{VAR\_POP}, \texttt{STDDEV\_SAMP}, \texttt{STDDEV\_POP} \\
2D Stats & \texttt{COVAR\_SAMP}, \texttt{COVAR\_POP}, \texttt{CORR}, \\
 & \texttt{REGR\_COUNT}, \texttt{REGR\_AVGX}, \texttt{REGR\_AVGY}, \\
 & \texttt{REGR\_SYY}, \texttt{REGR\_SXX}, \texttt{REGR\_SXY}, \\
 & \texttt{REGR\_SLOPE}, \texttt{REGR\_INTERCEPT}, \texttt{REGR\_R2}
\end{tabular}

\vspace{-6pt}
\caption{Supported aggregate functions for pre-aggregated views.}
\label{tbl:supported-aggs}
\end{table}

We next check that $q_v$ uses only the supported aggregate functions shown in Table~\ref{tbl:supported-aggs}.
Among these functions, aggregates with a \texttt{FILTER} clause are supported, while those that include a \texttt{DISTINCT} clause are \emph{not} supported.
In principle, our model can also support sketch-based approximations: a sketch (such as HyperLogLog \cite{doi:10.46298/dmtcs.3545} to approximate \texttt{COUNT(DISTINCT)} or t-digest \cite{tdigest} for quantile estimation) could be constructed as a measure value, and merged in subsequent selection queries to the materialized view.
Unfortunately, common SQL variants do not include constructs to work with sketches as a first-class data type.
We return to this point in our future work discussion (\S\ref{sec:future-work}).

\subsection{Materialized View Dimensions}
\label{sec:opt-groupby}

To determine the dimensions of the pre-aggregated materialized view, we first extract all columns referenced in the \texttt{GROUP BY} clause of $q_v$.
We then add the dimensions corresponding to the active clause $c_a$.
For a \emph{point}-type clause, we simply bin by the discrete point values in each selected column.
For example, if a clause selects $(A = a_i) \texttt{\;AND\;} (B = b_j)$, we include the values of columns $A$ and $B$ as view dimensions.
For an \emph{interval}-type clause, we determine an appropriate pixel-level binning scheme based on the provided \emph{scales}, \emph{bin}, and \emph{pixelSize} metadata of $c_a$, as described in \S\ref{sec:sel-meta}.
When repeatedly querying for updates, the exact values provided for the $c_a$-derived dimensions (the \emph{point} values or \emph{interval} endpoints) will change in response to user input.

\subsection{Materialized View Measures}
\label{sec:opt-stats}

Materialized view measure columns consist of \emph{sufficient statistics}: pre-aggregated data, binned by the dimensions, from which the requested aggregates can be constructed.
For example, to construct a \texttt{COUNT} aggregate, we can apply the \texttt{COUNT} function upon materialization, then query the materialized view for updates using the \texttt{SUM} function.
Functions such as \texttt{SUM}, \texttt{PRODUCT}, \texttt{MIN}, \texttt{MAX}, and logical aggregates can be applied for both construction and update queries.
In these cases, only a single measure column is needed.

\begin{figure}[t]
\centering
\includegraphics[width=\linewidth]{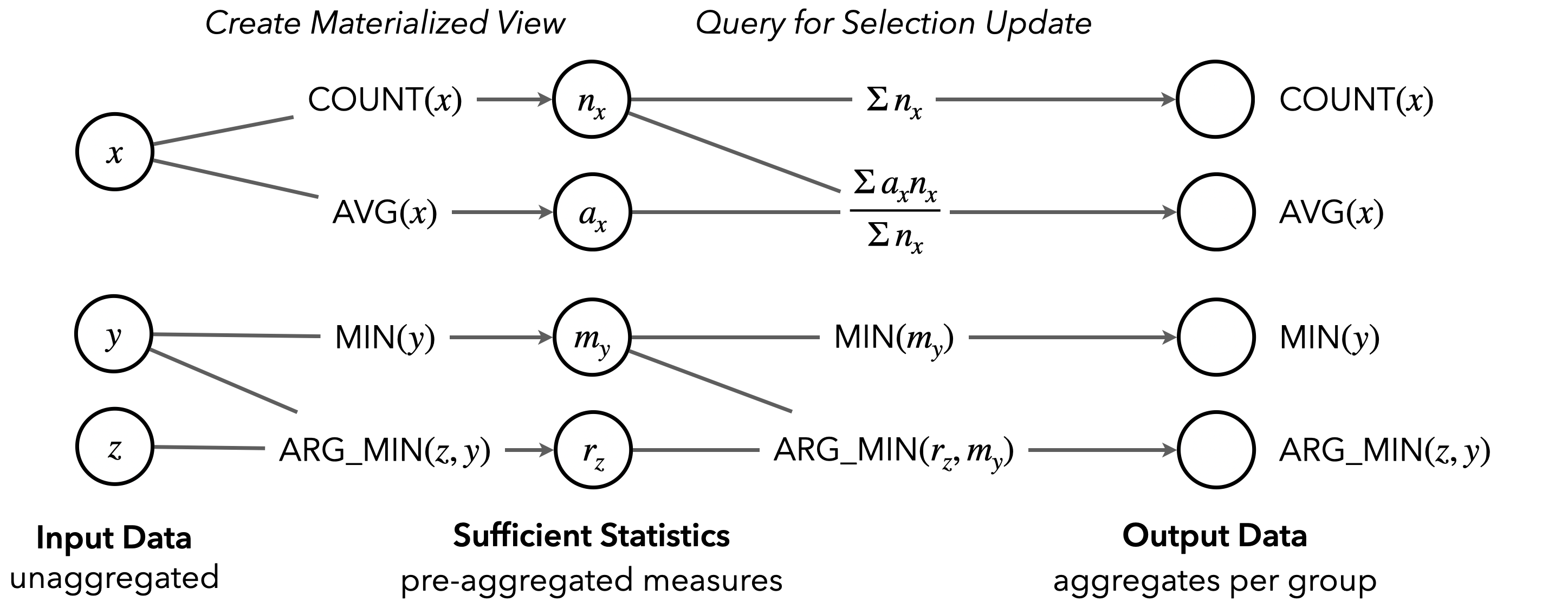}\vspace{-6pt}
\caption{Pre-aggregation and querying for univariate measures. Each sufficient statistic is included as a column in a materialized view, alongside grouping dimensions.}
\label{fig:materialized-view-uni}
\end{figure}

Other aggregates may require two or more measure columns.
An \texttt{AVG} (average) requires both \texttt{AVG} and \texttt{COUNT} statistics, from which a weighted average can be calculated.
For \texttt{ARG\_MIN}, both the \texttt{ARG\_MIN} and \texttt{MIN} values are needed (and similarly for \texttt{ARG\_MAX}).
These constructions are illustrated in Figure~\ref{fig:materialized-view-uni}.

\begin{figure}[t]
\centering
\includegraphics[width=\linewidth]{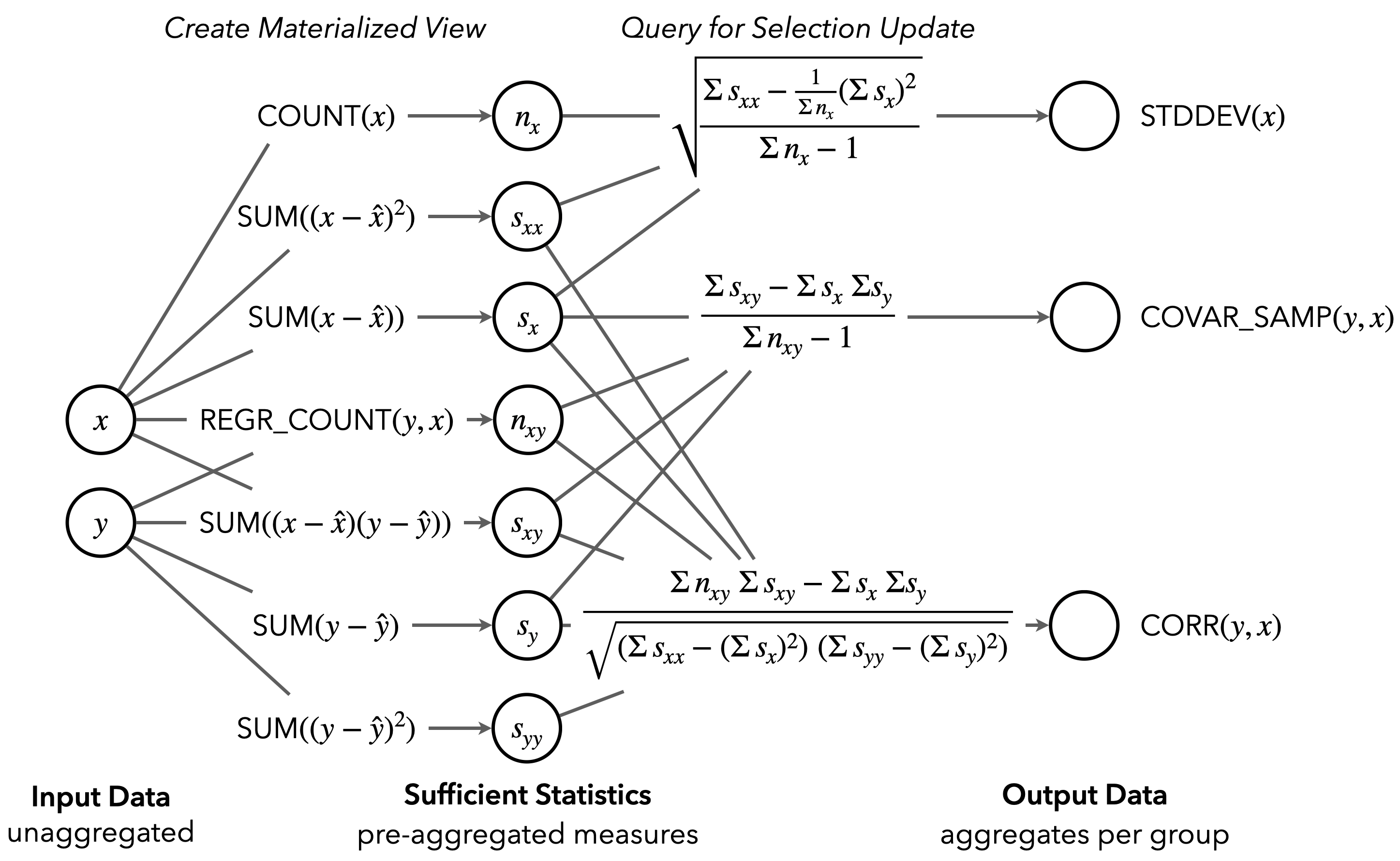}\vspace{-6pt}
\caption{Pre-aggregation and querying for standard deviation and bivariate measures. Each sufficient statistic is included as a column in a materialized view, alongside grouping dimensions. The symbol $\hat{x}$ indicates the average value of $x$ across the full dataset; it is included to mean-center the data to prevent floating point error.}\vspace{-12pt}
\label{fig:materialized-view-bi}
\end{figure}

Variance statistics such as \texttt{STDEV\_SAMP} and bivariate statistics for covariance, correlation, and linear regression can also be optimized.
Figure~\ref{fig:materialized-view-bi} shows the construction and update query steps for some of these aggregates.
For variance statistics, we must account for potential catastrophic cancellation due to floating point error.
Accordingly, we mean-center the data by first subtracting the global mean for a column \cite{doi:10.1080/00031305.1983.10483115}, denoted by $\hat{x}$ and $\hat{y}$ in Figure~\ref{fig:materialized-view-bi}.
Global means are computed using a scalar subquery, which retains linear-time computational complexity for the overall construction query.

\subsection{Querying Materialized Views}
\label{sec:querying-materialized-views}

Once created, materialized views can be queried to provide selection updates.
To do so, the current \emph{point} values or \emph{interval} endpoints in the active clause must be mapped to dimension bins, then applied in a \texttt{WHERE} clause to filter the pre-aggregated materialized view to just the selected cells.
Appropriate aggregation functions combine the sufficient statistics in the measure columns to produce output values.
This process is applied in the update query of Figure~\ref{fig:pre-agg-sql}, while various update aggregations are presented in Figure~\ref{fig:materialized-view-uni} and Figure~\ref{fig:materialized-view-bi}.

\subsection{Materialized View Representation}
\label{sec:mat-view-rep}

We represent a materialized view as a standard database table, with columns for groupby dimensions and pre-aggregated measures.
This representation has multiple benefits: the tables are sparse (they do not include rows for ``empty'' cells) and updates can be calculated in-database using standard aggregation queries.
In contrast, prior work such as imMens \cite{doi:10.1111/cgf.12129} and Falcon \cite{doi:10.1145/3290605.3300924} represent pre-aggregates (sometimes referred to as ``data tiles'') as dense multidimensional arrays.
These systems require a separate update query mechanism for these arrays, and their dense representation can result in arrays that do not fit in memory when a sparse representation may be tractable.
While pre-aggregates in Falcon use summed area tables \cite{doi:10.1145/800031.808600} to support constant time summation, in practice we have found querying materialized views to be sufficiently performant.
Moreover, these prior systems support only \texttt{COUNT} and \texttt{SUM} aggregates, as opposed to the richer set of aggregate functions described here.

Another advantage of a table-based representation is that pre-aggregated data can be stored and reused in-database.
To facilitate reuse, each table is named using a hash of its creation query string.
For example, lines 2--7 of the creation query in Figure~\ref{fig:pre-agg-sql} are hashed to determine the table name.
Future selection updates---whether in the same session or from another user accessing the same database---will only construct a materialized view if it does not already exist.

The selection model supports both on-demand materialized view generation and precomputation.
For dataset sizes up to \textasciitilde{}100M rows, on-demand materialized view generation is feasible (\S\ref{sec:eval}).
At the high end these may incur a \textasciitilde{}1 second delay for materialized view construction, but followed by real-time responses to active clause updates.
As discussed later, our implementation also performs prefetching to reduce materialization latency.
On-demand generation supports ad-hoc querying using multiple selection clauses: new materialized views are constructed for an active clause $c_a$ when one or more other selection clauses have changed.
As more clauses lead to greater selectivity, materialization becomes faster as users apply multiple clauses in tandem.

For datasets with a billion+ rows it is valuable to precompute materialized views.
For each interactor, materialized views for their clause structures may be precomputed at compile time, assuming these clauses are applied in isolation, and not intersected with other clauses.
If on-demand generation is too costly, an application can resolve selections using \texttt{LAST} aggregation to limit selections to a single clause only, ensuring the precomputed materialized views are sufficient.

\section{Implementation in Mosaic}
\label{sec:impl}

We implemented our selection model (\S\ref{sec:sel}) and optimizations (\S\ref{sec:opt}) in Mosaic \cite{doi:10.1109/TVCG.2023.3327189}, a software architecture for linking databases and interactive views.
Mosaic applications consist of a set of client views implemented in JavaScript that represent their data needs as SQL queries.
Each client view may be associated with a \emph{selection} for filtering the view's data.
Each view $v$ is responsible for incorporating a resolved selection predicate $p$ into its query ($v.query(p) \rightarrow q_{vp}$).

A central \emph{coordinator} is responsible for analyzing, optimizing, and issuing queries.
The system uses DuckDB \cite{doi:10.1145/3299869.3320212} as the backing database engine, though in principle other database systems could be used.
DuckDB provides a balance of performance and deployment flexibility.
Mosaic applications may run using DuckDB in the browser (via WebAssembly) \cite{doi:10.14778/3554821.3554847}, as a local or remote server, or within a Jupyter notebook kernel (including access to local Python data frames).

The binary Apache Arrow format is used to transfer data from the database to the JavaScript-based coordinator, enabling efficient data serialization with minimal subsequent parsing overhead.

The Mosaic coordinator supports optimizations such as query caching, query consolidation (merging compatible queries into a single issued query), and prefetching (client views may speculatively generate queries before results are needed).
Critically, the coordinator also applies the methods of \S\ref{sec:opt} to automatically materialize pre-aggregates to optimize selection updates.
For more details about the overall Mosaic architecture---including a library of \emph{view} and \emph{interactor} components spanning input widgets, tables, and interactive visualizations---see \cite{doi:10.1109/TVCG.2023.3327189}.
Here we focus on selection implementation and event handling details.

\subsection{Selection Lifecycle}
\label{sec:selection-lifecycle}

\emph{Selections} are reactive variables that manage clause sets, support predicate resolution, and dispatch update events to subscribed components.

\subsubsection{Selection Creation}
\label{sec:selection-creation}

A new selection instance $S = \langle C, R \rangle$ is created with an empty clause set ($C = \emptyset$) and parameters that define the resolution operator $R$.
These parameters are the aggregation method (e.g., \texttt{INTERSECT}, \texttt{UNION}, \texttt{LAST}), whether or not to apply cross-filtering (\emph{cross}), and whether an empty selection should select all records (the default) or none (\emph{empty}).
Once created, $S$ may be associated with one or more client views ($v \in V$), indicating that the views should be dynamically filtered by the resolved predicate provided by $R$.
At present, a client view $v$ may be associated with one and only one selection $S$.
The Mosaic coordinator subscribes for notifications when $S$ updates.

Selections can also be composed.
A selection may incorporate the clause sets of upstream selections via the \emph{includes} parameter.
Upon receiving a new clause, upstream selections relay that clause for inclusion in downstream selections, which then apply their own resolution operator $R$ to the combined clause set.
This composition mechanism permits fine-grained updates.
For example, a drop-down menu might filter autocomplete options for a text input (via an initial selection), while together the widgets filter a visualization (via a downstream selection).

\subsubsection{Selection Update and Event Dispatch}
\label{sec:selection-update-and-event-dispatch}

An interactor (\emph{i}) generates a new clause $c_i$ and passes it to $S$.
Any prior clause from the interactor \emph{i} is removed from $C$, then $c_i$ is added:

\vspace{-6pt}\[C \leftarrow (C \setminus \lbrace c \in C \;|\; source(c) = i \rbrace ) \cup \lbrace c_i \rbrace\]The clause $c_i$ is marked as the \emph{active} clause, and a selection \emph{update} event is queued for dispatch.
To provide rate limiting across asynchronous updates, the selection's event dispatcher tracks if a prior event is still being processed and waits before issuing a new event if so.
Update event handlers can return a \texttt{Promise}, which represents the eventual completion (or failure) of an asynchronous operation.
The dispatcher waits for all such \texttt{Promise}s to resolve before issuing the next update.

This dispatch scheme decouples interactors from subsequent selection handling.
While the selection dispatcher waits for a prior update to complete, a user may continue to interact, for example by adjusting interval endpoints.
Such incoming updates may supersede a prior update pending in the dispatch queue.
In these cases, the dispatcher will drop a previously queued update (in effect ignoring that input) such that only the most recent update is dispatched.
This throttling ensures smoother and more performant updates across input events.

\subsubsection{Selection Application}
\label{sec:selection-application}

When the coordinator receives a selection update from $S$, it must request new data for each affected view $v$ filtered by $S$.
The coordinator first checks if the (unfiltered) view query $q_v$ and active clause $c_a$ are amenable to pre-aggregation (\S\ref{sec:opt-compat}).
If so, a materialized view is created (if it hasn't already) and queried using the current values of $c_a$.
If pre-aggregation is not applicable, the coordinator applies the resolution operator $R(C, v)$ to produce a view-specific predicate $p$.
Given the predicate $p$, the view $v$ produces a filtered query $q_{vp}$, which the coordinator then issues to the database.
Once an update query completes, the coordinator returns the result (or error) to the client view $v$.

\subsubsection{Selection Activation}
\label{sec:selection-activation}

In addition to selection updates, interactors can enable prefetching by signaling when an update \emph{might} occur, providing an example selection clause prior to any actual user-driven selections.
For example, when a pointer (e.g., mouse cursor) enters an interactive region, this provides evidence that a user may soon interact with that component \cite{doi:10.1145/3290605.3300924}.
In such cases an interactor $i$ can pass an example clause to $S$, providing sufficient information for construction of a materialized view.
The selection $S$ then dispatches an \emph{activate} event, and the coordinator checks compatibility and initiates construction of a materialized view.
Mosaic interactors currently trigger activation upon pointer enter, at times providing prefetching lead times of a second or more.
In the future, other prediction models (cf. ForeCache \cite{doi:10.1145/2882903.2882919}) could be applied.

\subsection{Selection Scope}
\label{sec:selection-scope}

Mosaic focuses on filtering data using predicates provided to a SQL \texttt{WHERE} clause, which pushes filtering prior to other operations.
SQL variants provide additional filtering operations.
Aggregate operations may include operation-specific \texttt{FILTER} clauses; such clauses can pass through Mosaic's pre-aggregation optimization, so long as they do not include dynamic selection predicates.
A \texttt{QUALIFY} clause filters results involving window operations.
Though Mosaic does not apply selections in \texttt{QUALIFY} clauses, windowed values may be computed in a subquery, then filtered by a \texttt{WHERE} clause.
Finally, \texttt{HAVING} clauses filter results after aggregate operations.
Selection clauses could be applied to aggregate results (e.g., an interval selection in a continuous color legend of aggregated values); however, we do not concern ourselves with optimizing this case as data reduction has already been performed.

\section{Performance Benchmarks}
\label{sec:eval}

\begin{figure}[t]
\centering
\includegraphics[width=\linewidth]{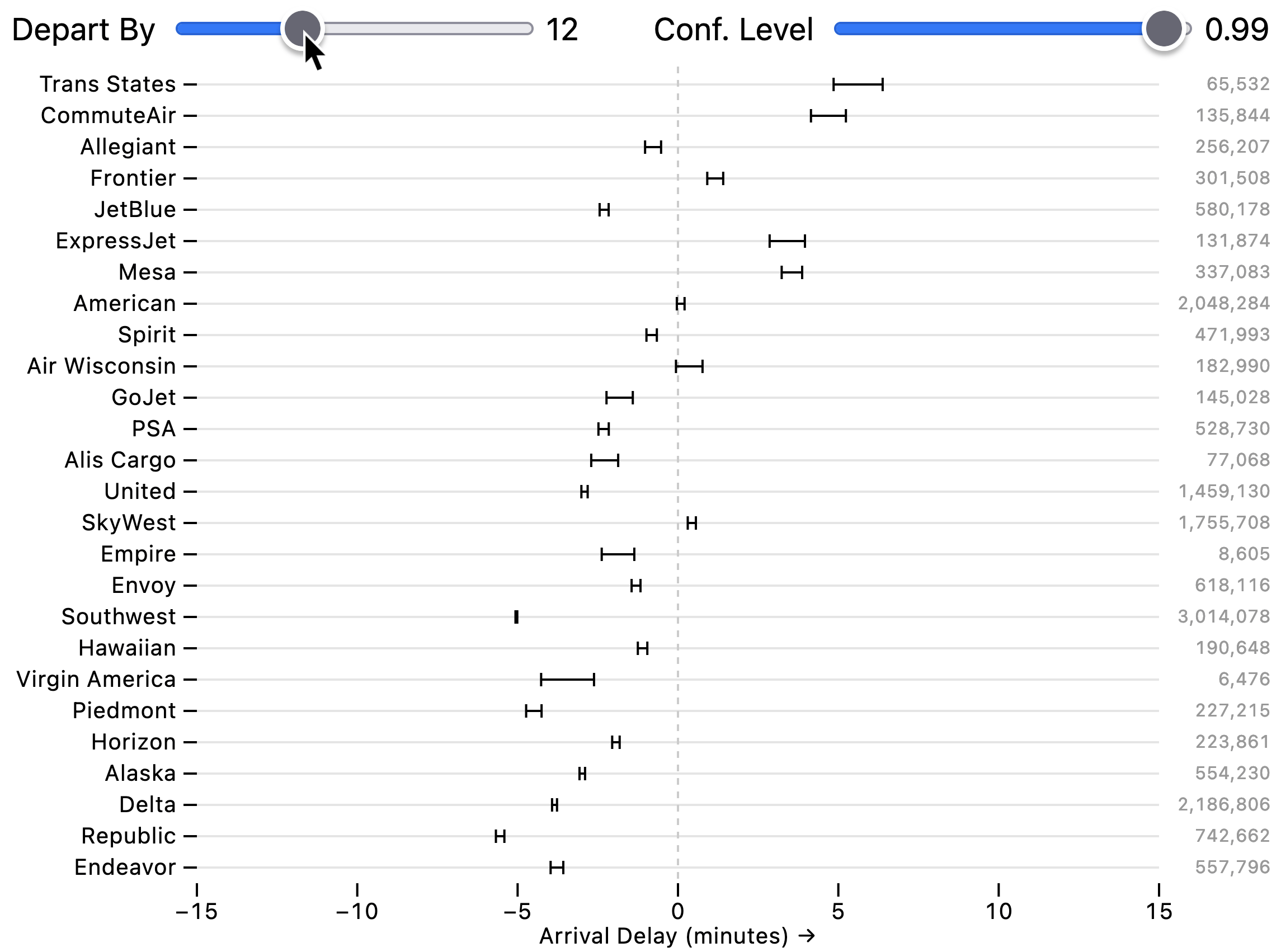}\vspace{-6pt}
\caption{Flight delay confidence intervals by airline. A slider filters the data to all flights leaving by the selected time.}\vspace{-6pt}
\label{fig:example-airlines}
\end{figure}

\begin{figure}[t]
\centering
\includegraphics[width=\linewidth]{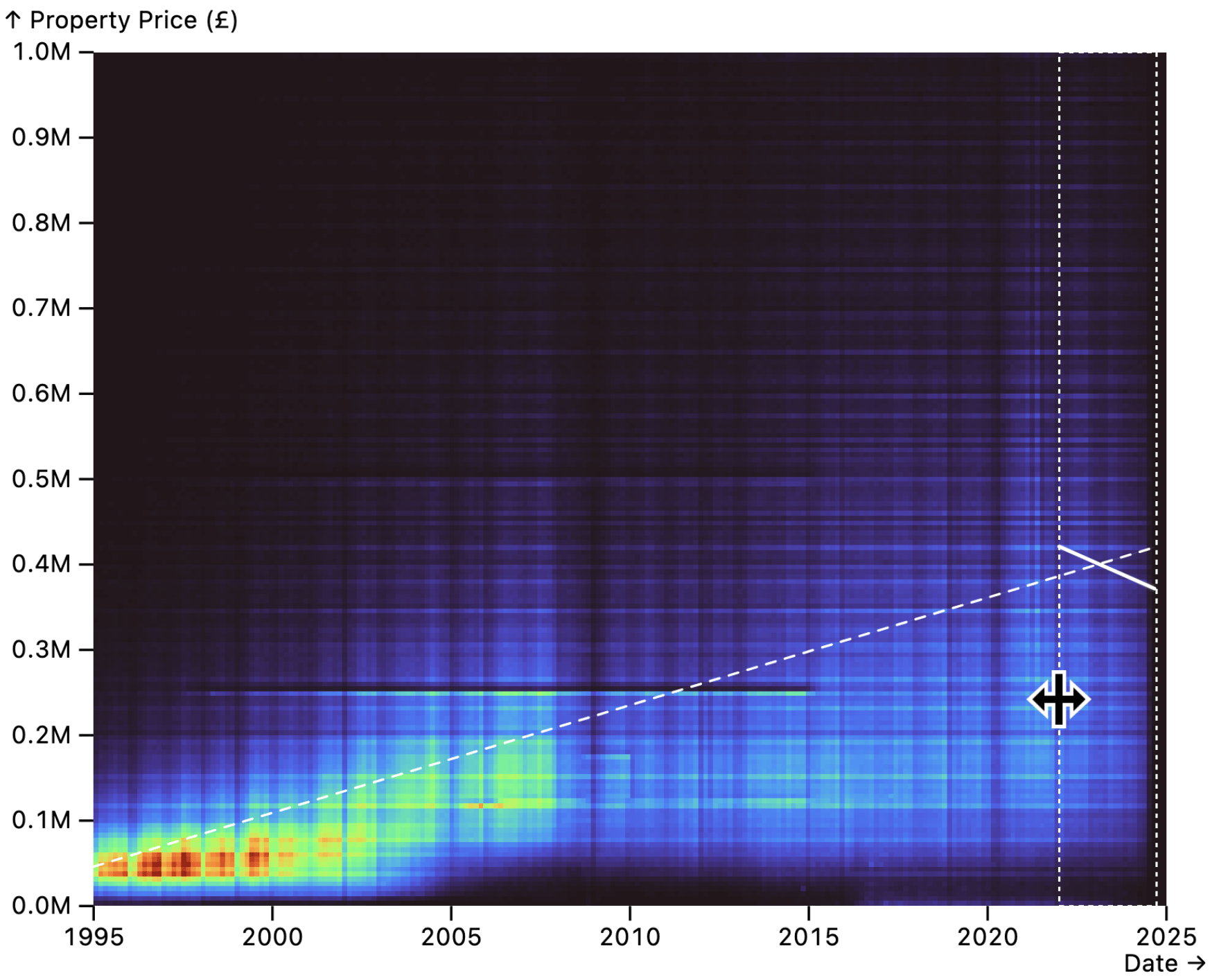}
\vspace{-14pt}\vspace{-6pt}
\caption{Density plot and regression fits for United Kingdom property prices by date. A local regression is fit within the selected interval, revealing post-Brexit price declines.}\vspace{-12pt}
\label{fig:example-property}
\end{figure}

Given a wealth of empirical work establishing the benefits of reduced latency \cite{psych-hci, doi:10.1037/1076-898X.6.4.322, doi:10.1109/TVCG.2014.2346452, doi:10.1109/TVCG.2016.2607714}, we focus our evaluation on query performance.
We conduct performance benchmarks to evaluate Mosaic selections and pre-aggregation optimizations, measuring both materialized view creation and selection update query times over a collection of visualization templates.
These include the flight histograms (Figure~\ref{fig:example-flights}, plus a third histogram for flight distance), NYC taxi (Figure~\ref{fig:example-taxi}), and Gaia sky survey (Figure~\ref{fig:example-gaia}) examples shown earlier.
The Gaia and NYC taxi examples update high-resolution rasters, requiring larger materialized views.
Additionally, we include (1) flight delay confidence intervals by airline (Figure~\ref{fig:example-airlines}), filtered by latest departure, to test average and standard deviation aggregates; and (2) a density plot of date vs. sales price for 29.5M properties in the United Kingdom (Figure~\ref{fig:example-property}), to test interactive regression fits over a selected interval.
All visualizations require multiple update queries upon selection changes, except for the U.K. property example, which only updates the regression fit.

For each visualization template we simulate user selections for all interactive components.
For 1D and 2D interval selections, we use intervals that are 10\%, 20\%, and 30\% of the axis range, and sweep them across the display.
For slider inputs (airline delay interval example), we sweep through all positions.
The resulting queries for materialized view creation and selection updates form our benchmark suite.

We vary the base table size from 10 thousand to 1 billion records.
As the flights and property datasets have fewer than 100M rows, for these datasets we upsample the data to create the 100 million (100M) and 1 billion (1B) row conditions.
We sample with replacement while adding random noise to numeric columns.

We run tests against optimized Mosaic selections in both DuckDB and DuckDB-WASM (WebAssembly) \cite{doi:10.14778/3554821.3554847}.
DuckDB-WASM is deployable in browsers, limited to 4GB of memory (due to 32-bit addressing) and is single-threaded by default.
We use DuckDB v1.0 and DuckDB-WASM v1.29.0 with default configurations.
As a baseline, we include unoptimized selections that issue queries against a base table rather than creating and querying materialized views.
We ran the benchmarks within Node.js v20.11.1 on a 2021 MacBook Pro laptop (16-inch, MacOS 14.5) with an M1 Pro processor and 16GB RAM.

For comparison with other state-of-the-art visualization tools that perform automatic optimizations, we include Yang et al.~\cite{doi:10.1145/3639276}'s results for VegaFusion \cite{doi:10.1109/VIS54862.2022.00011} and VegaPlus \cite{doi:10.1145/3639276} on the flight histograms example, as they used the same visualization design and comparable setup.
Note that neither VegaFusion nor VegaPlus optimize raster visualizations as found in the U.K. property, Gaia, and NYC taxi examples.

\begin{figure*}[t]
\centering
\includegraphics[width=\linewidth]{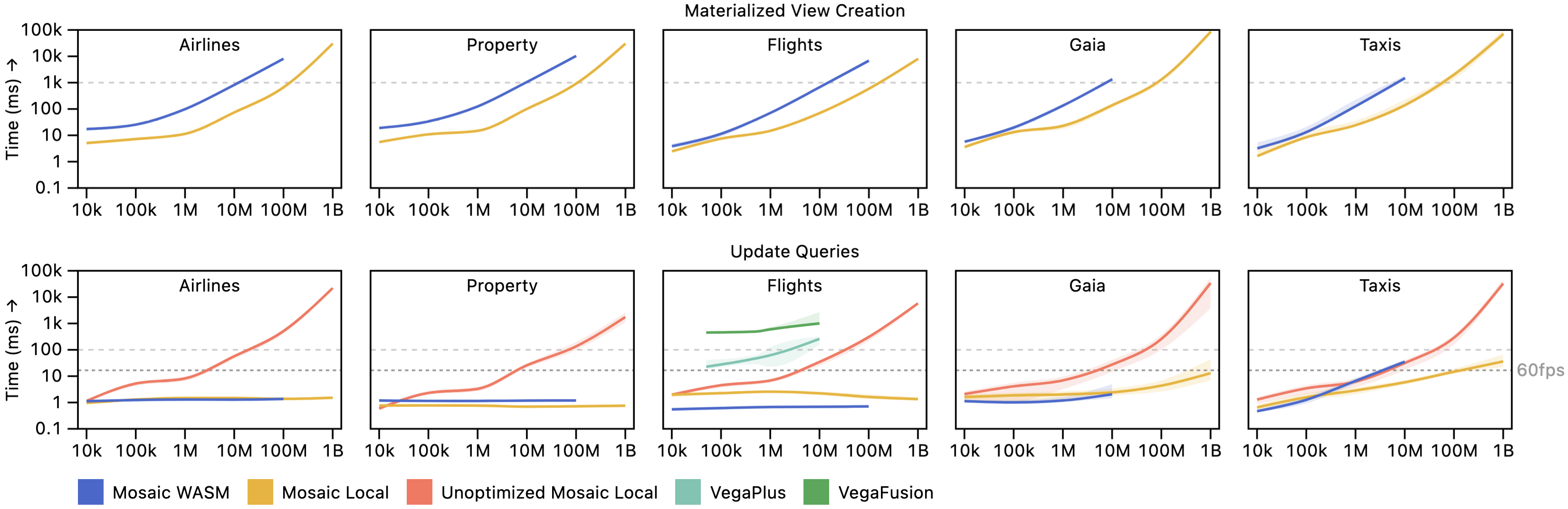}\vspace{-6pt}
\caption{Performance benchmark results for pre-aggregated materialized view construction and selection update queries. Lines show median query completion times in milliseconds, while shaded areas indicate interquartile ranges. Both axes are base 10 log-transformed. As dataset sizes increase, optimized selections preserve low-latency updates (typically 60fps or faster), while requiring at most 1 second for materialized view creation for dataset sizes up to 100M records. To frontload longer (\textgreater{} 1s) construction times, precomputation may be combined with \texttt{LAST} clause resolution (\S\ref{sec:mat-view-rep}).}\vspace{-12pt}
\label{fig:benchmark-results}
\end{figure*}

Figure~\ref{fig:benchmark-results} plots the benchmark results.
Up to dataset sizes of 100M, materialized view creation on a local DuckDB instance completes within 1 second or less, supporting ``cold start'' exploration.
For larger dataset sizes with 1B rows, view creation takes 1--2 minutes, slow enough to make compile-time precomputation attractive.
As a consequence of being single-threaded, DuckDB-WASM is predictably slower, creating materialized views within 1 second or less for datasets of 10 million or fewer rows.
Furthermore, due to WebAssembly limitations DuckDB-WASM runs out of memory when attempting to load the largest datasets (1B rows, and 100M row Gaia and NYC Taxi datasets).

Using pre-aggregated materialized views, Mosaic selections in DuckDB and DuckDB-WASM provide performant updates under the 100ms threshold \cite{psych-hci}, and typically much faster (1--10ms).
Unoptimized Mosaic selections degrade as a function of dataset size, becoming orders of magnitude slower and crossing the 100ms threshold for larger dataset sizes.
For the flight histograms, VegaFusion and VegaPlus provide slower responses than unoptimized Mosaic selections.
This result suggests that the Mosaic selection model's direct application to SQL queries is more efficient, as other tools perform cross-platform transformations from a JavaScript runtime.

For all but one visualization template, materialized views provide faster results than unoptimized update queries at all dataset sizes.
Nevertheless, the results can suggest places where pre-aggregation optimization may be unnecessary.
Unoptimized queries for the U.K. property visualization provide slightly faster results at the smallest dataset size, as calculating regression fits from a materialized view requires many sufficient statistics (see Figure~\ref{fig:materialized-view-bi}), introducing additional overhead.
However, at the largest dataset sizes, pre-aggregation for regression fits provides faster responses by 3--4 orders of magnitude.

Beyond latency, the size of materialized views is an important concern.
Appendix B analyzes materialized view sizes, demonstrating smaller, denser views for standard charts like histograms and larger, sparser views involving higher-resolution raster plots.

\section{Future Work and Limitations}
\label{sec:future-work}

While Mosaic selections are sufficient to express and optimize a variety of application configurations, they have limitations.
First, pre-aggregation optimizations currently assume a static dataset. Our approach for dynamic datasets is to delete and then reconstruct materialized views.
Future work might apply incremental updates instead.

Next, Mosaic selections apply predicates to a single source relation.
While the same selection can be applied to different tables, the schemas (column names and types) must overlap to ensure valid predicates.
In addition, multiple tables can be joined and then filtered.
One compelling case is network data, where edges might be filtered based on the attributes of connected nodes.
In such cases, join results can be materialized as a new table or expressed as a database view.
Queries over large joins might be supported using lightweight materialization \cite{doi:10.1145/3626735}.

Pre-aggregation assumes a stable binning domain.
Dynamic binning is supported, but requires recomputation of pre-aggregated materialized views.
If dynamic binning is intermittent (e.g., due to pan/zoom operations), subsequent brushing can be optimized between re-binning operations.
If dynamic binning is a direct result of brushing (and the client \texttt{filterStable} property is false), Mosaic gracefully degrades by issuing standard queries without pre-aggregation.

As our performance benchmarks indicate, DuckDB-WASM does not scale past 100M records.
Such larger datasets are supported using a sufficiently resourced database server.
Here we tested this using server-side DuckDB, leveraging parallelism on a single machine.
Future work might examine distributed data engines for even larger datasets.
To handle more extreme data volumes, Mosaic supports the use of sampling for rapid exploration, which can be followed by longer running full queries to validate results (c.f., Moritz et al.~\cite{doi:10.1145/3025453.3025456}).

Partitioning work across local and remote databases may provide additional benefits \cite{moritz2015dynamic}.
A remote database server might construct pre-aggregated materialized views, then send them to a local database instance (e.g., DuckDB-WASM) for faster updates without network latency.
One route is for the Mosaic architecture to explicitly manage such partitions, another is for database systems to do so automatically (cf. MotherDuck \cite{atwal2024motherduck}).
Further study is needed to determine optimized partitions with or without leveraging application-level information.

Many other avenues remain for future work.
The Mosaic selection model is in principle compatible with sketches (\S\ref{sec:opt-compat}): as sketches can be merged, they could serve as sufficient statistics for approximate cardinality and quantile estimation.
However, while many databases provide sketch-based aggregates, they typically do not expose the sketches as first-class data types.
Adding such data types and corresponding merge operations would enable additional optimizations.

Alternatively, databases might add internal support for selection-driven materialized views.
Mosaic's pre-aggregated materialized views share similarities with \texttt{PROJECTION} clauses, a form of explicit materialized view found in database systems such as Vertica and ClickHouse.
However, a more powerful model might be to provide an update query template---similar to a prepared statement, but for projections---where the selection values (such as interval endpoints) are included as bounded parameters.
The database engine might then internally determine sufficient statistics, support even more aggregate functions (including sketches), and perform incremental view maintenance \cite{doi:10.7551/mitpress/4472.003.0016}.

Other directions include further optimizations enabled by the selection model.
Currently, client views in Mosaic can perform prefetching (e.g., for paging data while scrolling a table or panning and zooming a visualization), but they are responsible for issuing the prefetch queries and stitching query results together.
The selection model might instead enable such prefetching optimizations \emph{automatically}.
For example, panning and zooming can be expressed as changes to a 1D or 2D interval selection, which determine the data extents of the current viewport.
Given a client view query, view screen size information, and selection metadata, an optimizer might automatically determine an appropriate tiling scheme, perform prefetching (e.g., for adjacent tiles), and stitch together tiles to provide rapid updates as a user navigates.
Appendix A presents benchmark results demonstrating the benefits of both data presorting and prefetching for scalable panning interactions.

\section{Conclusion}
\label{sec:conclusion}

We presented Mosaic selections---a general model of user-driven selections in interactive data systems---alongside optimization techniques that analyze an active selection clause and client view queries to automatically create and query pre-aggregated materialized views.
Queries to materialized views are then applied in tandem with other optimizations including visualization-specific queries (e.g., M4 \cite{doi:10.14778/2732951.2732953, kohn2023dashql}), query caching, and query consolidation.
Mosaic selections provide a practical solution to translate between interface and database layers, optimizing workloads of many rapid queries in interactive data systems.
Selection information simplifies the creation and querying of pre-aggregated materialized views, sidestepping the difficulties of general approaches to multi-query optimization and answering queries with views.

Performance benchmarks demonstrate that these optimizations provide low-latency interactive updates (under 100ms, and typically 1--10ms) while supporting ``cold start'' exploration by building materialized views on the fly for datasets with up to hundreds of millions of rows.
For larger datasets, pre-aggregated materialized views may be created prior to visual analysis.
Meanwhile, both unoptimized Mosaic queries and existing automatic visualization optimization methods (VegaFusion and VegaPlus) are significantly slower, providing unacceptably long latencies on datasets with millions or more rows.

The Mosaic selection model and optimizations are available as open source software at \href{https://idl.uw.edu/mosaic}{idl.uw.edu/mosaic}. Benchmark code and results are available at \href{https://github.com/uwdata/mosaic-selection-benchmarks}{github.com/uwdata/mosaic-selection-benchmarks}.

\acknowledgments{We thank the UW Interactive Data Lab and CMU DIG Lab for their feedback. This work was supported by NSF award IIS 2402718.}

\bibliographystyle{abbrv-doi-hyperref}
\bibliography{index.bib}

\begin{thebibliography}{10}

\bibitem{atwal2024motherduck}
R.~Atwal, P.~A. Boncz, R.~Boyd, A.~Courtney, T.~D{\" o}hmen, F.~Gerlinghoff,
  J.~Huang, J.~Hwang, R.~Hyde, E.~Felder, and {others}.
\newblock {MotherDuck}: {DuckDB} in the cloud and in the client.
\newblock In {\em Proc. Conference on Innovative Data Research ({CIDR})}, 2024.

\bibitem{doi:10.1145/2882903.2882919}
L.~Battle, R.~Chang, and M.~Stonebraker.
\newblock Dynamic {Prefetching} of {Data} {Tiles} for {Interactive}
  {Visualization}.
\newblock In {\em Proc. {ACM} {Conference} on {Management} of {Data} (SIGMOD)},
  pp. 1363--1375. ACM, 2016. \href{https://doi.org/10.1145/2882903.2882919}
{doi: {{%
10\hspace{.1pt}\discretionary{.}{%
}{.}\hspace{.4pt}1145\discretionary{/}{%
}{/}2882903\hspace{.1pt}\discretionary{.}{%
}{.}\hspace{.4pt}2882919}}}


\bibitem{doi:10.1111/cgf.13678}
L.~Battle and J.~Heer.
\newblock Characterizing {Exploratory} {Visual} {Analysis}: A {Literature}
  {Review} and {Evaluation} of {Analytic} {Provenance} in {Tableau}.
\newblock {\em Computer Graphics Forum (Proc. EuroVis)}, 38(3):145--159, 2019.
  \href{https://doi.org/10.1111/cgf.13678}
{doi: {{%
10\hspace{.1pt}\discretionary{.}{%
}{.}\hspace{.4pt}1111\discretionary{/}{%
}{/}cgf\hspace{.1pt}\discretionary{.}{%
}{.}\hspace{.4pt}13678}}}


\bibitem{doi:10.1109/TVCG.2020.3028891}
L.~Battle and C.~Scheidegger.
\newblock A {Structured} {Review} of {Data} {Management} {Technology} for
  {Interactive} {Visualization} and {Analysis}.
\newblock {\em IEEE Transactions on Visualization and Computer Graphics},
  27(2):1128--1138, 2021. \href{https://doi.org/10.1109/tvcg.2020.3028891}
{doi: {{%
10\hspace{.1pt}\discretionary{.}{%
}{.}\hspace{.4pt}1109\discretionary{/}{%
}{/}tvcg\hspace{.1pt}\discretionary{.}{%
}{.}\hspace{.4pt}2020\hspace{.1pt}\discretionary{.}{%
}{.}\hspace{.4pt}3028891}}}


\bibitem{doi:10.1080/00401706.1987.10488204}
R.~A. Becker and W.~S. Cleveland.
\newblock Brushing {Scatterplots}.
\newblock {\em Technometrics}, 29(2):127--142, 1987.
  \href{https://doi.org/10.1080/00401706.1987.10488204}
{doi: {{%
10\hspace{.1pt}\discretionary{.}{%
}{.}\hspace{.4pt}1080\discretionary{/}{%
}{/}00401706\hspace{.1pt}\discretionary{.}{%
}{.}\hspace{.4pt}1987\hspace{.1pt}\discretionary{.}{%
}{.}\hspace{.4pt}10488204}}}


\bibitem{doi:10.1109/TVCG.2011.185}
M.~Bostock, V.~Ogievetsky, and J.~Heer.
\newblock D3 {Data}-{Driven} {Documents}.
\newblock {\em IEEE Transactions on Visualization and Computer Graphics},
  17(12):2301--2309, 2011. \href{https://doi.org/10.1109/tvcg.2011.185}
{doi: {{%
10\hspace{.1pt}\discretionary{.}{%
}{.}\hspace{.4pt}1109\discretionary{/}{%
}{/}tvcg\hspace{.1pt}\discretionary{.}{%
}{.}\hspace{.4pt}2011\hspace{.1pt}\discretionary{.}{%
}{.}\hspace{.4pt}185}}}


\bibitem{bts-ontime}
{Bureau of Transportation Statistics}.
\newblock On-{Time} {Performance}.
\newblock \href{https://www.bts.gov/}{https://www.bts.gov/}, 2023.

\bibitem{psych-hci}
S.~K. Card, T.~P. Moran, and A.~Newell.
\newblock {\em The {Psychology} of {Human}-{Computer} {Interaction}}.
\newblock L. Erlbaum Associates Inc., 1983.

\bibitem{doi:10.1080/00031305.1983.10483115}
T.~F. Chan, G.~H. Golub, and R.~J. Leveque.
\newblock Algorithms for {Computing} the {Sample} {Variance}: Analysis and
  {Recommendations}.
\newblock {\em The American Statistician}, 37(3):242--247, 1983.
  \href{https://doi.org/10.1080/00031305.1983.10483115}
{doi: {{%
10\hspace{.1pt}\discretionary{.}{%
}{.}\hspace{.4pt}1080\discretionary{/}{%
}{/}00031305\hspace{.1pt}\discretionary{.}{%
}{.}\hspace{.4pt}1983\hspace{.1pt}\discretionary{.}{%
}{.}\hspace{.4pt}10483115}}}


\bibitem{hong2003compound}
H.~Chen.
\newblock Compound brushing.
\newblock In {\em Proc. {IEEE} {Information} {Visualization}}, pp. 181--188.
  IEEE Computer Society, Seattle, Washington, 2003.
  \href{https://doi.org/10.5555/1947368.1947402}
{doi: {{%
10\hspace{.1pt}\discretionary{.}{%
}{.}\hspace{.4pt}5555\discretionary{/}{%
}{/}1947368\hspace{.1pt}\discretionary{.}{%
}{.}\hspace{.4pt}1947402}}}


\bibitem{doi:10.1145/800031.808600}
F.~C. Crow.
\newblock Summed-area tables for texture mapping.
\newblock In {\em Proc. {Computer} Graphics and Interactive Techniques
  ({SIGGRAPH})}, pp. 207--212. ACM, 1984.
  \href{https://doi.org/10.1145/800031.808600}
{doi: {{%
10\hspace{.1pt}\discretionary{.}{%
}{.}\hspace{.4pt}1145\discretionary{/}{%
}{/}800031\hspace{.1pt}\discretionary{.}{%
}{.}\hspace{.4pt}808600}}}


\bibitem{doi:10.1145/263407.263545}
M.~Derthick, J.~Kolojejchick, and S.~F. Roth.
\newblock An interactive visual query environment for exploring data.
\newblock In {\em Proc. {ACM} Symposium on {User} Interface Software and
  Technology ({UIST})}, pp. 189--198. ACM Press, 1997.
  \href{https://doi.org/10.1145/263407.263545}
{doi: {{%
10\hspace{.1pt}\discretionary{.}{%
}{.}\hspace{.4pt}1145\discretionary{/}{%
}{/}263407\hspace{.1pt}\discretionary{.}{%
}{.}\hspace{.4pt}263545}}}


\bibitem{tdigest}
T.~Dunning.
\newblock The t-digest: Efficient estimates of distributions.
\newblock {\em Software Impacts}, 7, 2021.
  \href{https://doi.org/10.1016/j.simpa.2020.100049}
{doi: {{%
10\hspace{.1pt}\discretionary{.}{%
}{.}\hspace{.4pt}1016\discretionary{/}{%
}{/}j\hspace{.1pt}\discretionary{.}{%
}{.}\hspace{.4pt}simpa\hspace{.1pt}\discretionary{.}{%
}{.}\hspace{.4pt}2020\hspace{.1pt}\discretionary{.}{%
}{.}\hspace{.4pt}100049}}}


\bibitem{doi:10.46298/dmtcs.3545}
P.~Flajolet, {\' E}.~Fusy, O.~Gandouet, and F.~Meunier.
\newblock Hyperloglog: the analysis of a near-optimal cardinality estimation
  algorithm.
\newblock {\em Discrete Mathematics \& Theoretical Computer Science}, DMTCS
  Proceedings vol. AH,..., 2007. \href{https://doi.org/10.46298/dmtcs.3545}
{doi: {{%
10\hspace{.1pt}\discretionary{.}{%
}{.}\hspace{.4pt}46298\discretionary{/}{%
}{/}dmtcs\hspace{.1pt}\discretionary{.}{%
}{.}\hspace{.4pt}3545}}}


\bibitem{gaia-release}
{Gaia Collaboration}, {Smart, R. L.}, {Sarro, L. M.}, and {422 others}.
\newblock Gaia {Early} {Data} {Release} 3 - {The} {Gaia} {Catalogue} of
  {Nearby} {Stars}.
\newblock {\em Astronomy and Astrophysics}, 649, 2021.
  \href{https://doi.org/10.1051/0004-6361/202039498}
{doi: {{%
10\hspace{.1pt}\discretionary{.}{%
}{.}\hspace{.4pt}1051\discretionary{/}{%
}{/}0004\discretionary{%
}{-}{-}6361\discretionary{/}{%
}{/}202039498}}}


\bibitem{gaia:gdr3}
{Gaia Collaboration}, A.~Vallenari, A.~G.~A. Brown, and {453 others}.
\newblock Gaia {Data} {Release} 3: Summary of the content and survey
  properties.
\newblock 2022. \href{https://doi.org/10.48550/arXiv.2208.00211}
{doi: {{%
10\hspace{.1pt}\discretionary{.}{%
}{.}\hspace{.4pt}48550\discretionary{/}{%
}{/}arXiv\hspace{.1pt}\discretionary{.}{%
}{.}\hspace{.4pt}2208\hspace{.1pt}\discretionary{.}{%
}{.}\hspace{.4pt}00211}}}


\bibitem{datacube}
J.~Gray, S.~Chaudhuri, A.~Bosworth, A.~Layman, D.~Reichart, M.~Venkatrao,
  F.~Pellow, and H.~Pirahesh.
\newblock Data cube: A relational aggregation operator generalizing group-by,
  cross-tab, and sub-totals.
\newblock {\em Data Mining and Knowledge Discovery}, 1:29--53, 1997.

\bibitem{doi:10.1037/1076-898X.6.4.322}
W.~D. Gray and D.~A. Boehm-Davis.
\newblock Milliseconds matter: An introduction to microstrategies and to their
  use in describing and predicting interactive behavior.
\newblock {\em Journal of Experimental Psychology: Applied}, 6(4):322--335,
  2000. \href{https://doi.org/10.1037/1076-898x.6.4.322}
{doi: {{%
10\hspace{.1pt}\discretionary{.}{%
}{.}\hspace{.4pt}1037\discretionary{/}{%
}{/}1076\discretionary{%
}{-}{-}898x\hspace{.1pt}\discretionary{.}{%
}{.}\hspace{.4pt}6\hspace{.1pt}\discretionary{.}{%
}{.}\hspace{.4pt}4\hspace{.1pt}\discretionary{.}{%
}{.}\hspace{.4pt}322}}}


\bibitem{doi:10.7551/mitpress/4472.003.0016}
A.~Gupta and I.~S. Mumick.
\newblock {\em Maintenance of {Materialized} {Views}: Problems, {Techniques},
  and {Applications}}, pp. 145--158.
\newblock The MIT Press, 1999.
  \href{https://doi.org/10.7551/mitpress/4472.003.0016}
{doi: {{%
10\hspace{.1pt}\discretionary{.}{%
}{.}\hspace{.4pt}7551\discretionary{/}{%
}{/}mitpress\discretionary{/}{%
}{/}4472\hspace{.1pt}\discretionary{.}{%
}{.}\hspace{.4pt}003\hspace{.1pt}\discretionary{.}{%
}{.}\hspace{.4pt}0016}}}


\bibitem{doi:10.1007/s007780100054}
A.~Y. Halevy.
\newblock Answering queries using views: A survey.
\newblock {\em The VLDB Journal}, 10(4):270--294, 2001.
  \href{https://doi.org/10.1007/s007780100054}
{doi: {{%
10\hspace{.1pt}\discretionary{.}{%
}{.}\hspace{.4pt}1007\discretionary{/}{%
}{/}s007780100054}}}


\bibitem{doi:10.1145/235968.233333}
V.~Harinarayan, A.~Rajaraman, and J.~D. Ullman.
\newblock Implementing data cubes efficiently.
\newblock {\em ACM SIGMOD Record}, 25(2):205--216, 1996.
  \href{https://doi.org/10.1145/235968.233333}
{doi: {{%
10\hspace{.1pt}\discretionary{.}{%
}{.}\hspace{.4pt}1145\discretionary{/}{%
}{/}235968\hspace{.1pt}\discretionary{.}{%
}{.}\hspace{.4pt}233333}}}


\bibitem{doi:10.1109/VIS49827.2021.9623323}
J.~Heer.
\newblock Fast \& {Accurate} {Gaussian} {Kernel} {Density} {Estimation}.
\newblock In {\em 2021 {IEEE} {Visualization} {Conference} ({VIS})}, pp.
  11--15. IEEE, 2021. \href{https://doi.org/10.1109/vis49827.2021.9623323}
{doi: {{%
10\hspace{.1pt}\discretionary{.}{%
}{.}\hspace{.4pt}1109\discretionary{/}{%
}{/}vis49827\hspace{.1pt}\discretionary{.}{%
}{.}\hspace{.4pt}2021\hspace{.1pt}\discretionary{.}{%
}{.}\hspace{.4pt}9623323}}}


\bibitem{doi:10.1145/1357054.1357203}
J.~Heer, M.~Agrawala, and W.~Willett.
\newblock Generalized selection via interactive query relaxation.
\newblock In {\em Proc. {SIGCHI} {Conference} on {Human} {Factors} in
  {Computing} {Systems} (CHI)}. ACM, 2008.
  \href{https://doi.org/10.1145/1357054.1357203}
{doi: {{%
10\hspace{.1pt}\discretionary{.}{%
}{.}\hspace{.4pt}1145\discretionary{/}{%
}{/}1357054\hspace{.1pt}\discretionary{.}{%
}{.}\hspace{.4pt}1357203}}}


\bibitem{doi:10.1109/TVCG.2023.3327189}
J.~Heer and D.~Moritz.
\newblock Mosaic: An {Architecture} for {Scalable} \& {Interoperable} {Data}
  {Views}.
\newblock {\em IEEE Transactions on Visualization and Computer Graphics}, pp.
  1--11, 2023. \href{https://doi.org/10.1109/tvcg.2023.3327189}
{doi: {{%
10\hspace{.1pt}\discretionary{.}{%
}{.}\hspace{.4pt}1109\discretionary{/}{%
}{/}tvcg\hspace{.1pt}\discretionary{.}{%
}{.}\hspace{.4pt}2023\hspace{.1pt}\discretionary{.}{%
}{.}\hspace{.4pt}3327189}}}


\bibitem{doi:10.1145/2133806.2133821}
J.~Heer and B.~Shneiderman.
\newblock Interactive dynamics for visual analysis.
\newblock {\em Communications of the ACM}, 55(4):45--54, 2012.
  \href{https://doi.org/10.1145/2133806.2133821}
{doi: {{%
10\hspace{.1pt}\discretionary{.}{%
}{.}\hspace{.4pt}1145\discretionary{/}{%
}{/}2133806\hspace{.1pt}\discretionary{.}{%
}{.}\hspace{.4pt}2133821}}}


\bibitem{doi:10.1145/3626735}
Z.~Huang and E.~Wu.
\newblock Lightweight {Materialization} for {Fast} {Dashboards} {Over} {Joins}.
\newblock {\em Proc. {ACM} {Conference} on {Management} of {Data} (SIGMOD)},
  1(4):1--27, 2023. \href{https://doi.org/10.1145/3626735}
{doi: {{%
10\hspace{.1pt}\discretionary{.}{%
}{.}\hspace{.4pt}1145\discretionary{/}{%
}{/}3626735}}}


\bibitem{doi:10.1080/00949658308810650}
M.~C. Jones and H.~W. Lotwick.
\newblock On the errors involved in computing the empirical characteristic
  function.
\newblock {\em Journal of Statistical Computation and Simulation},
  17(2):133--149, 1983. \href{https://doi.org/10.1080/00949658308810650}
{doi: {{%
10\hspace{.1pt}\discretionary{.}{%
}{.}\hspace{.4pt}1080\discretionary{/}{%
}{/}00949658308810650}}}


\bibitem{doi:10.14778/2732951.2732953}
U.~Jugel, Z.~Jerzak, G.~Hackenbroich, and V.~Markl.
\newblock M4.
\newblock {\em Proc. VLDB Endowment}, 7(10):797--808, 2014.
  \href{https://doi.org/10.14778/2732951.2732953}
{doi: {{%
10\hspace{.1pt}\discretionary{.}{%
}{.}\hspace{.4pt}14778\discretionary{/}{%
}{/}2732951\hspace{.1pt}\discretionary{.}{%
}{.}\hspace{.4pt}2732953}}}


\bibitem{doi:10.1145/3034786.3034792}
T.~Kathuria and S.~Sudarshan.
\newblock Efficient and {Provable} {Multi}-{Query} {Optimization}.
\newblock In {\em Proc. {ACM} {SIGMOD}-{SIGACT}-{SIGAI} {Symposium} on
  {Principles} of {Database} {Systems}}, pp. 53--67. ACM, 2017.
  \href{https://doi.org/10.1145/3034786.3034792}
{doi: {{%
10\hspace{.1pt}\discretionary{.}{%
}{.}\hspace{.4pt}1145\discretionary{/}{%
}{/}3034786\hspace{.1pt}\discretionary{.}{%
}{.}\hspace{.4pt}3034792}}}


\bibitem{kohn2023dashql}
A.~Kohn, D.~Moritz, and T.~Neumann.
\newblock {DashQL} -- {Complete} {Analysis} {Workflows} with {SQL}.
\newblock 2023. \href{https://doi.org/10.48550/ARXIV.2306.03714}
{doi: {{%
10\hspace{.1pt}\discretionary{.}{%
}{.}\hspace{.4pt}48550\discretionary{/}{%
}{/}ARXIV\hspace{.1pt}\discretionary{.}{%
}{.}\hspace{.4pt}2306\hspace{.1pt}\discretionary{.}{%
}{.}\hspace{.4pt}03714}}}


\bibitem{doi:10.14778/3554821.3554847}
A.~Kohn, D.~Moritz, M.~Raasveldt, H.~M{\" u}hleisen, and T.~Neumann.
\newblock {DuckDB-WASM}.
\newblock {\em Proc. VLDB Endowment}, 15(12):3574--3577, 2022.
  \href{https://doi.org/10.14778/3554821.3554847}
{doi: {{%
10\hspace{.1pt}\discretionary{.}{%
}{.}\hspace{.4pt}14778\discretionary{/}{%
}{/}3554821\hspace{.1pt}\discretionary{.}{%
}{.}\hspace{.4pt}3554847}}}


\bibitem{doi:10.1109/VIS54862.2022.00011}
N.~Kruchten, J.~Mease, and D.~Moritz.
\newblock {VegaFusion}: Automatic {Server}-{Side} {Scaling} for {Interactive}
  {Vega} {Visualizations}.
\newblock In {\em Proc. {IEEE} {VIS} Short Papers}, pp. 11--15. IEEE, 2022.
  \href{https://doi.org/10.1109/vis54862.2022.00011}
{doi: {{%
10\hspace{.1pt}\discretionary{.}{%
}{.}\hspace{.4pt}1109\discretionary{/}{%
}{/}vis54862\hspace{.1pt}\discretionary{.}{%
}{.}\hspace{.4pt}2022\hspace{.1pt}\discretionary{.}{%
}{.}\hspace{.4pt}00011}}}


\bibitem{doi:10.1109/TVCG.2013.179}
L.~Lins, J.~T. Klosowski, and C.~Scheidegger.
\newblock Nanocubes for {Real}-{Time} {Exploration} of {Spatiotemporal}
  {Datasets}.
\newblock {\em IEEE Transactions on Visualization and Computer Graphics},
  19(12):2456--2465, 2013. \href{https://doi.org/10.1109/tvcg.2013.179}
{doi: {{%
10\hspace{.1pt}\discretionary{.}{%
}{.}\hspace{.4pt}1109\discretionary{/}{%
}{/}tvcg\hspace{.1pt}\discretionary{.}{%
}{.}\hspace{.4pt}2013\hspace{.1pt}\discretionary{.}{%
}{.}\hspace{.4pt}179}}}


\bibitem{doi:10.1109/TVCG.2014.2346452}
Z.~Liu and J.~Heer.
\newblock The {Effects} of {Interactive} {Latency} on {Exploratory} {Visual}
  {Analysis}.
\newblock {\em IEEE Transactions on Visualization and Computer Graphics},
  20(12):2122--2131, 2014. \href{https://doi.org/10.1109/tvcg.2014.2346452}
{doi: {{%
10\hspace{.1pt}\discretionary{.}{%
}{.}\hspace{.4pt}1109\discretionary{/}{%
}{/}tvcg\hspace{.1pt}\discretionary{.}{%
}{.}\hspace{.4pt}2014\hspace{.1pt}\discretionary{.}{%
}{.}\hspace{.4pt}2346452}}}


\bibitem{doi:10.1111/cgf.12129}
Z.~Liu, B.~Jiang, and J.~Heer.
\newblock \textit{{imMens}}: Realtime {Visual} {Querying} of {Big} {Data}.
\newblock {\em Computer Graphics Forum (Proc. EuroVis)}, 32(3pt4):421--430,
  2013. \href{https://doi.org/10.1111/cgf.12129}
{doi: {{%
10\hspace{.1pt}\discretionary{.}{%
}{.}\hspace{.4pt}1111\discretionary{/}{%
}{/}cgf\hspace{.1pt}\discretionary{.}{%
}{.}\hspace{.4pt}12129}}}


\bibitem{doi:10.1145/253262.253335}
M.~Livny, R.~Ramakrishnan, K.~Beyer, G.~Chen, D.~Donjerkovic, S.~Lawande,
  J.~Myllymaki, and K.~Wenger.
\newblock Devise.
\newblock {\em ACM SIGMOD Record}, 26(2):301--312, 1997.
  \href{https://doi.org/10.1145/253262.253335}
{doi: {{%
10\hspace{.1pt}\discretionary{.}{%
}{.}\hspace{.4pt}1145\discretionary{/}{%
}{/}253262\hspace{.1pt}\discretionary{.}{%
}{.}\hspace{.4pt}253335}}}


\bibitem{doi:10.14778/3407790.3407826}
H.~Mohammed, Z.~Wei, E.~Wu, and R.~Netravali.
\newblock Continuous prefetch for interactive data applications.
\newblock {\em Proc. VLDB Endowment}, 13(12):2297--2311, 2020.
  \href{https://doi.org/10.14778/3407790.3407826}
{doi: {{%
10\hspace{.1pt}\discretionary{.}{%
}{.}\hspace{.4pt}14778\discretionary{/}{%
}{/}3407790\hspace{.1pt}\discretionary{.}{%
}{.}\hspace{.4pt}3407826}}}


\bibitem{doi:10.48550/arXiv.1808.06019}
D.~Moritz and D.~Fisher.
\newblock Visualizing a {Million} {Time} {Series} with the {Density} {Line}
  {Chart}.
\newblock 2018. \href{https://doi.org/10.48550/ARXIV.1808.06019}
{doi: {{%
10\hspace{.1pt}\discretionary{.}{%
}{.}\hspace{.4pt}48550\discretionary{/}{%
}{/}ARXIV\hspace{.1pt}\discretionary{.}{%
}{.}\hspace{.4pt}1808\hspace{.1pt}\discretionary{.}{%
}{.}\hspace{.4pt}06019}}}


\bibitem{doi:10.1145/3025453.3025456}
D.~Moritz, D.~Fisher, B.~Ding, and C.~Wang.
\newblock Trust, but {Verify}: Optimistic {Visualizations} of {Approximate}
  {Queries} for {Exploring} {Big} {Data}.
\newblock In {\em Proc. ACM {Conference} on {Human} {Factors} in {Computing}
  {Systems} (CHI)}, pp. 2904--2915. ACM, 2017.
  \href{https://doi.org/10.1145/3025453.3025456}
{doi: {{%
10\hspace{.1pt}\discretionary{.}{%
}{.}\hspace{.4pt}1145\discretionary{/}{%
}{/}3025453\hspace{.1pt}\discretionary{.}{%
}{.}\hspace{.4pt}3025456}}}


\bibitem{moritz2015dynamic}
D.~Moritz, J.~Heer, and B.~Howe.
\newblock Dynamic client-server optimization for scalable interactive
  visualization on the web.
\newblock In {\em Workshop on {Data} {Systems} for {Interactive} {Analysis}
  ({DSIA})}, 2015.

\bibitem{doi:10.1145/3290605.3300924}
D.~Moritz, B.~Howe, and J.~Heer.
\newblock Falcon: Balancing interactive latency and resolution sensitivity for
  scalable linked visualizations.
\newblock In {\em Proc. ACM {Conference} on {Human} {Factors} in {Computing}
  {Systems} (CHI)}, pp. 1--11. ACM, 2019.
  \href{https://doi.org/10.1145/3290605.3300924}
{doi: {{%
10\hspace{.1pt}\discretionary{.}{%
}{.}\hspace{.4pt}1145\discretionary{/}{%
}{/}3290605\hspace{.1pt}\discretionary{.}{%
}{.}\hspace{.4pt}3300924}}}


\bibitem{doi:10.1145/345513.345282}
C.~North and B.~Shneiderman.
\newblock Snap-together visualization: A user interface for coordinating
  visualizations via relational schemata.
\newblock In {\em Proc. Conference on {Advanced} Visual Interfaces ({AVI})},
  pp. 128--135. ACM, 2000. \href{https://doi.org/10.1145/345513.345282}
{doi: {{%
10\hspace{.1pt}\discretionary{.}{%
}{.}\hspace{.4pt}1145\discretionary{/}{%
}{/}345513\hspace{.1pt}\discretionary{.}{%
}{.}\hspace{.4pt}345282}}}


\bibitem{olston1998viqing}
C.~Olston, M.~Stonebraker, A.~Aiken, and J.~M. Hellerstein.
\newblock {VIQING}: Visual {Interactive} {QueryING}.
\newblock In {\em Proc. {IEEE} {Symposium} on {Visual} {Languages}}, p. 162.
  IEEE Computer Society, USA, 1998.
  \href{https://doi.org/10.5555/832279.834493}
{doi: {{%
10\hspace{.1pt}\discretionary{.}{%
}{.}\hspace{.4pt}5555\discretionary{/}{%
}{/}832279\hspace{.1pt}\discretionary{.}{%
}{.}\hspace{.4pt}834493}}}


\bibitem{doi:10.1145/3299869.3320212}
M.~Raasveldt and H.~M{\" u}hleisen.
\newblock {DuckDB}: An embedded analytical database.
\newblock In {\em Proc. {ACM} {Conference} on {Management} of {Data} (SIGMOD)}.
  ACM, 2019. \href{https://doi.org/10.1145/3299869.3320212}
{doi: {{%
10\hspace{.1pt}\discretionary{.}{%
}{.}\hspace{.4pt}1145\discretionary{/}{%
}{/}3299869\hspace{.1pt}\discretionary{.}{%
}{.}\hspace{.4pt}3320212}}}


\bibitem{doi:10.1145/342009.335419}
P.~Roy, S.~Seshadri, S.~Sudarshan, and S.~Bhobe.
\newblock Efficient and extensible algorithms for multi query optimization.
\newblock In {\em Proc. {ACM} {Conference} on {Management} of {Data} (SIGMOD)}.
  ACM, 2000. \href{https://doi.org/10.1145/342009.335419}
{doi: {{%
10\hspace{.1pt}\discretionary{.}{%
}{.}\hspace{.4pt}1145\discretionary{/}{%
}{/}342009\hspace{.1pt}\discretionary{.}{%
}{.}\hspace{.4pt}335419}}}


\bibitem{doi:10.1109/TVCG.2016.2599030}
A.~Satyanarayan, D.~Moritz, K.~Wongsuphasawat, and J.~Heer.
\newblock Vega-{Lite}: A {Grammar} of {Interactive} {Graphics}.
\newblock {\em IEEE Transactions on Visualization and Computer Graphics},
  23(1):341--350, 2017. \href{https://doi.org/10.1109/tvcg.2016.2599030}
{doi: {{%
10\hspace{.1pt}\discretionary{.}{%
}{.}\hspace{.4pt}1109\discretionary{/}{%
}{/}tvcg\hspace{.1pt}\discretionary{.}{%
}{.}\hspace{.4pt}2016\hspace{.1pt}\discretionary{.}{%
}{.}\hspace{.4pt}2599030}}}


\bibitem{doi:10.1109/TVCG.2015.2467091}
A.~Satyanarayan, R.~Russell, J.~Hoffswell, and J.~Heer.
\newblock Reactive {Vega}: A {Streaming} {Dataflow} {Architecture} for
  {Declarative} {Interactive} {Visualization}.
\newblock {\em IEEE Transactions on Visualization and Computer Graphics},
  22(1):659--668, 2016. \href{https://doi.org/10.1109/tvcg.2015.2467091}
{doi: {{%
10\hspace{.1pt}\discretionary{.}{%
}{.}\hspace{.4pt}1109\discretionary{/}{%
}{/}tvcg\hspace{.1pt}\discretionary{.}{%
}{.}\hspace{.4pt}2015\hspace{.1pt}\discretionary{.}{%
}{.}\hspace{.4pt}2467091}}}


\bibitem{doi:10.1145/42201.42203}
T.~K. Sellis.
\newblock Multiple-query optimization.
\newblock {\em ACM Transactions on Database Systems}, 13(1):23--52, 1988.
  \href{https://doi.org/10.1145/42201.42203}
{doi: {{%
10\hspace{.1pt}\discretionary{.}{%
}{.}\hspace{.4pt}1145\discretionary{/}{%
}{/}42201\hspace{.1pt}\discretionary{.}{%
}{.}\hspace{.4pt}42203}}}


\bibitem{doi:10.1109/2945.981851}
C.~Stolte, D.~Tang, and P.~Hanrahan.
\newblock Polaris: a system for query, analysis, and visualization of
  multidimensional relational databases.
\newblock {\em IEEE Transactions on Visualization and Computer Graphics},
  8(1):52--65, 2002. \href{https://doi.org/10.1109/2945.981851}
{doi: {{%
10\hspace{.1pt}\discretionary{.}{%
}{.}\hspace{.4pt}1109\discretionary{/}{%
}{/}2945\hspace{.1pt}\discretionary{.}{%
}{.}\hspace{.4pt}981851}}}


\bibitem{doi:10.1109/TVCG.2020.3030372}
W.~Tao, X.~Hou, A.~Sah, L.~Battle, R.~Chang, and M.~Stonebraker.
\newblock Kyrix-{S}: Authoring {Scalable} {Scatterplot} {Visualizations} of
  {Big} {Data}.
\newblock {\em IEEE Transactions on Visualization and Computer Graphics},
  27(2):401--411, 2021. \href{https://doi.org/10.1109/tvcg.2020.3030372}
{doi: {{%
10\hspace{.1pt}\discretionary{.}{%
}{.}\hspace{.4pt}1109\discretionary{/}{%
}{/}tvcg\hspace{.1pt}\discretionary{.}{%
}{.}\hspace{.4pt}2020\hspace{.1pt}\discretionary{.}{%
}{.}\hspace{.4pt}3030372}}}


\bibitem{doi:10.1111/cgf.13708}
W.~Tao, X.~Liu, Y.~Wang, L.~Battle, {\c C}.~Demiralp, R.~Chang, and
  M.~Stonebraker.
\newblock Kyrix: Interactive {Pan}/{Zoom} {Visualizations} at {Scale}.
\newblock {\em Computer Graphics Forum (Proc. EuroVis)}, 38(3):529--540, 2019.
  \href{https://doi.org/10.1111/cgf.13708}
{doi: {{%
10\hspace{.1pt}\discretionary{.}{%
}{.}\hspace{.4pt}1111\discretionary{/}{%
}{/}cgf\hspace{.1pt}\discretionary{.}{%
}{.}\hspace{.4pt}13708}}}


\bibitem{doi:10.1038/s41586-023-06415-8}
J.~L. Watson, D.~Juergens, N.~R. Bennett, B.~L. Trippe, J.~Yim, H.~E. Eisenach,
  W.~Ahern, A.~J. Borst, R.~J. Ragotte, L.~F. Milles, B.~I.~M. Wicky,
  N.~Hanikel, S.~J. Pellock, A.~Courbet, W.~Sheffler, J.~Wang, P.~Venkatesh,
  I.~Sappington, S.~V. Torres, A.~Lauko, V.~De~Bortoli, E.~Mathieu,
  S.~Ovchinnikov, R.~Barzilay, T.~S. Jaakkola, F.~DiMaio, M.~Baek, and
  D.~Baker.
\newblock De novo design of protein structure and function with {RFdiffusion}.
\newblock {\em Nature}, 620(7976):1089--1100, 2023.
  \href{https://doi.org/10.1038/s41586-023-06415-8}
{doi: {{%
10\hspace{.1pt}\discretionary{.}{%
}{.}\hspace{.4pt}1038\discretionary{/}{%
}{/}s41586\discretionary{%
}{-}{-}023\discretionary{%
}{-}{-}06415\discretionary{%
}{-}{-}8}}}


\bibitem{weaver2004improvise}
C.~Weaver.
\newblock Building {Highly}-{Coordinated} {Visualizations} in {Improvise}.
\newblock In {\em IEEE {Symposium} on {Information} {Visualization}}, pp.
  159--166, 2004. \href{https://doi.org/10.1109/INFVIS.2004.12}
{doi: {{%
10\hspace{.1pt}\discretionary{.}{%
}{.}\hspace{.4pt}1109\discretionary{/}{%
}{/}INFVIS\hspace{.1pt}\discretionary{.}{%
}{.}\hspace{.4pt}2004\hspace{.1pt}\discretionary{.}{%
}{.}\hspace{.4pt}12}}}


\bibitem{doi:10.1088/0957-0233/24/2/027001}
J.~B.~W. Webber.
\newblock A bi-symmetric log transformation for wide-range data.
\newblock {\em Measurement Science and Technology}, 24(2):027001, 2012.
  \href{https://doi.org/10.1088/0957-0233/24/2/027001}
{doi: {{%
10\hspace{.1pt}\discretionary{.}{%
}{.}\hspace{.4pt}1088\discretionary{/}{%
}{/}0957\discretionary{%
}{-}{-}0233\discretionary{/}{%
}{/}24\discretionary{/}{%
}{/}2\discretionary{/}{%
}{/}027001}}}


\bibitem{doi:10.1198/jcgs.2009.07098}
H.~Wickham.
\newblock A {Layered} {Grammar} of {Graphics}.
\newblock {\em Journal of Computational and Graphical Statistics}, 19(1):3--28,
  2010. \href{https://doi.org/10.1198/jcgs.2009.07098}
{doi: {{%
10\hspace{.1pt}\discretionary{.}{%
}{.}\hspace{.4pt}1198\discretionary{/}{%
}{/}jcgs\hspace{.1pt}\discretionary{.}{%
}{.}\hspace{.4pt}2009\hspace{.1pt}\discretionary{.}{%
}{.}\hspace{.4pt}07098}}}


\bibitem{doi:10.1007/978-3-642-21551-3_13}
L.~Wilkinson.
\newblock {\em The {Grammar} of {Graphics}}, pp. 375--414.
\newblock Springer Berlin Heidelberg, 2011.
  \href{https://doi.org/10.1007/978-3-642-21551-3_13}
{doi: {{%
10\hspace{.1pt}\discretionary{.}{%
}{.}\hspace{.4pt}1007\discretionary{/}{%
}{/}978\discretionary{%
}{-}{-}3\discretionary{%
}{-}{-}642\discretionary{%
}{-}{-}21551\discretionary{%
}{-}{-}3\_13}}}


\bibitem{doi:10.1109/TVCG.2021.3114796}
Y.~Wu, R.~Chang, J.~M. Hellerstein, A.~Satyanarayan, and E.~Wu.
\newblock {DIEL}: Interactive {Visualization} {Beyond} the {Here} and {Now}.
\newblock {\em IEEE Transactions on Visualization and Computer Graphics},
  28(1):737--746, 2022. \href{https://doi.org/10.1109/tvcg.2021.3114796}
{doi: {{%
10\hspace{.1pt}\discretionary{.}{%
}{.}\hspace{.4pt}1109\discretionary{/}{%
}{/}tvcg\hspace{.1pt}\discretionary{.}{%
}{.}\hspace{.4pt}2021\hspace{.1pt}\discretionary{.}{%
}{.}\hspace{.4pt}3114796}}}


\bibitem{doi:10.1145/3639276}
J.~Yang, H.~K. Joo, S.~Yerramreddy, D.~Moritz, and L.~Battle.
\newblock Optimizing {Dataflow} {Systems} for {Scalable} {Interactive}
  {Visualization}.
\newblock {\em Proc. {ACM} {Conference} on {Management} of {Data} (SIGMOD)},
  2(1):1--25, 2024. \href{https://doi.org/10.1145/3639276}
{doi: {{%
10\hspace{.1pt}\discretionary{.}{%
}{.}\hspace{.4pt}1145\discretionary{/}{%
}{/}3639276}}}


\bibitem{doi:10.1109/TVCG.2016.2607714}
E.~Zgraggen, A.~Galakatos, A.~Crotty, J.-D. Fekete, and T.~Kraska.
\newblock How {Progressive} {Visualizations} {Affect} {Exploratory} {Analysis}.
\newblock {\em IEEE Transactions on Visualization and Computer Graphics},
  23(8):1977--1987, 2017. \href{https://doi.org/10.1109/tvcg.2016.2607714}
{doi: {{%
10\hspace{.1pt}\discretionary{.}{%
}{.}\hspace{.4pt}1109\discretionary{/}{%
}{/}tvcg\hspace{.1pt}\discretionary{.}{%
}{.}\hspace{.4pt}2016\hspace{.1pt}\discretionary{.}{%
}{.}\hspace{.4pt}2607714}}}


\end{thebibliography}

\appendix
\clearpage\section{Data Sorting and Prefetching}
\label{sec:data-sorting-and-prefetching}

In addition to preaggregation, Mosaic selections can enable other forms of automatic optimization, either in real-time or via precomputation.
As a selection definition indicates the data access patterns an application will use, we can prepare the data to optimize such access paths, for example by pre-sorting the data along a filtering dimension.
Meanwhile, interactions such as pan and zoom involve predictable navigation patterns \cite{doi:10.1145/2882903.2882919}: we can prefetch data tiles that users are likely to visit.

\begin{figure}[t]
\centering
\includegraphics[width=\linewidth]{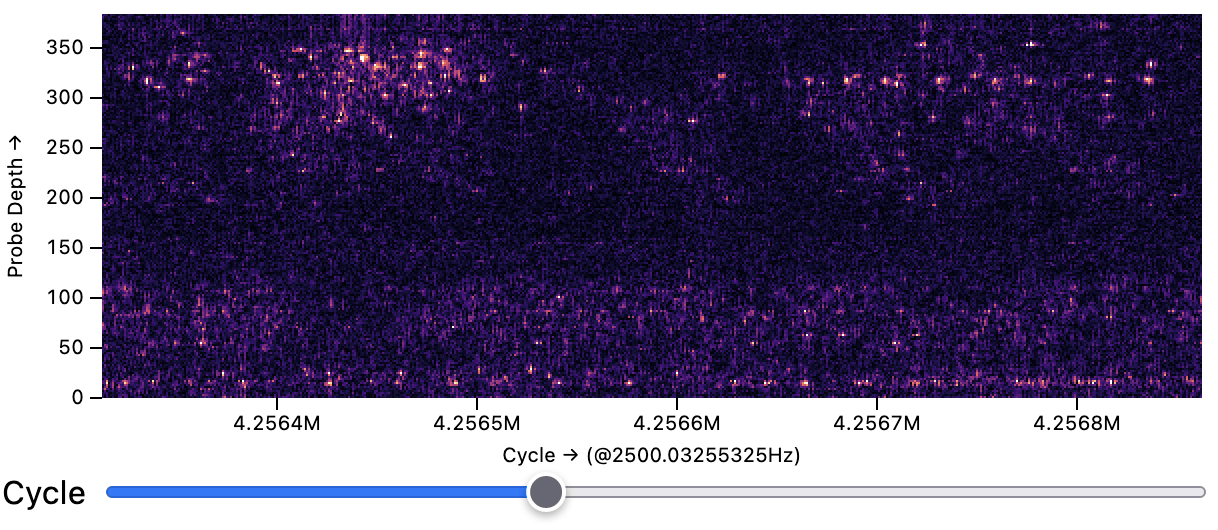}\vspace{-6pt}
\caption{Electrical recordings from a mouse brain, containing over 10.7M time samples (4.1B rows). The slider pans the display; an associated interval selection determines the viewport domain.}
\label{fig:example-neuro}
\end{figure}

The visualization in Figure~\ref{fig:example-neuro} depicts electrical recordings from neurons.
A probe with 384 recording channels was inserted into a mouse brain, with electrical activity sampled at \textasciitilde{}2500 Hz.
Roughly 71 minutes of recording results in 10.7M time samples for each channel, producing a table with over 4.1B rows.
Visualizing the raw recordings enables neuroscientists to assess data quality and validate results from neuron spike classifiers used to further analyze the data.
Figure~\ref{fig:example-neuro} visualizes the neuron recordings at the finest granularity as a raster display, while an interval selection determines the content of the viewport.
We desire low-latency updates when a user pans the display via scrolling or slider adjustments, updating the underlying selection interval.

\subsection{Sorting and Prefetching Optimizations}
\label{sec:sorting-and-prefetching-optimizations}

We test two different classes of optimization.
The first optimization is \emph{sorting} the data, a ``compile''-time optimization that can be performed before publishing an interactive data application.
The original data is ordered primarily by sensor depth (the 384 recording channels) and only secondarily by time. However, the panning interaction requires querying data along the time dimension.
By sorting the data to align with this access pattern, we should be able to reduce query times.
Access-path aligned sorting improves spatial locality and leverages underlying data representation features.
Databases such as DuckDB and file formats such as Parquet store data in blocks of rows along with minimum and maximum value statistics for each column.
These summary statistics enable a scanner to skip entire blocks if the selected interval and statistics do not overlap, preventing the need to actually read data values directly.

The second optimization is \emph{prefetching}: here, requesting adjacent data records (e.g., those immediately offscreen) before they are needed.
To effectively prefetch we first need a tiling scheme that divides the data into tiled regions that can be individually requested.
Then, queries for neighboring tiles might be issued prior to actual use.
We have implemented tiling in Mosaic using a specialized \texttt{raster} mark client view that uses the initial viewport domain as a tile size, queries the database on a per-tile basis, and then stitches requested tile query results together as needed.
This mark also calls the Mosaic coordinator's \texttt{prefetch} method to issue queries for adjacent tiles.
The coordinator runs these queries with lower priority and caches the results (using the same standard query caching mechanism) for later potential use.

\subsection{Benchmark Results}
\label{sec:benchmark-results}

We evaluate both sorting and prefetching optimizations over the neuron recording data.
We measure query latency upon panning interactions, in response to changes to the underlying interval selection.
To simulate user panning interactions, we alternate among \emph{step} interactions that pan the view by half the length of the current viewport (and thus can take advantage of adjacent prefetched tiles) and \emph{skip} interactions that jump to a different part of the data.
A session is modeled as 10 skips to unvisited regions of the data, each followed by 100 steps.

For sorting, we compare the original \emph{unsorted} data order (grouped by recording channel) to an optimized order \emph{sorted} by time.
For prefetching, we compare three conditions: \emph{direct} querying without use of tiles or prefetching, \emph{tile} queries without prefetching (to assess any overhead due to tiling), and \emph{prefetch} queries that perform tiling and also instruct the coordinator to prefetch immediately adjacent tiles.

\begin{figure}[t]
\centering
\includegraphics[width=\linewidth]{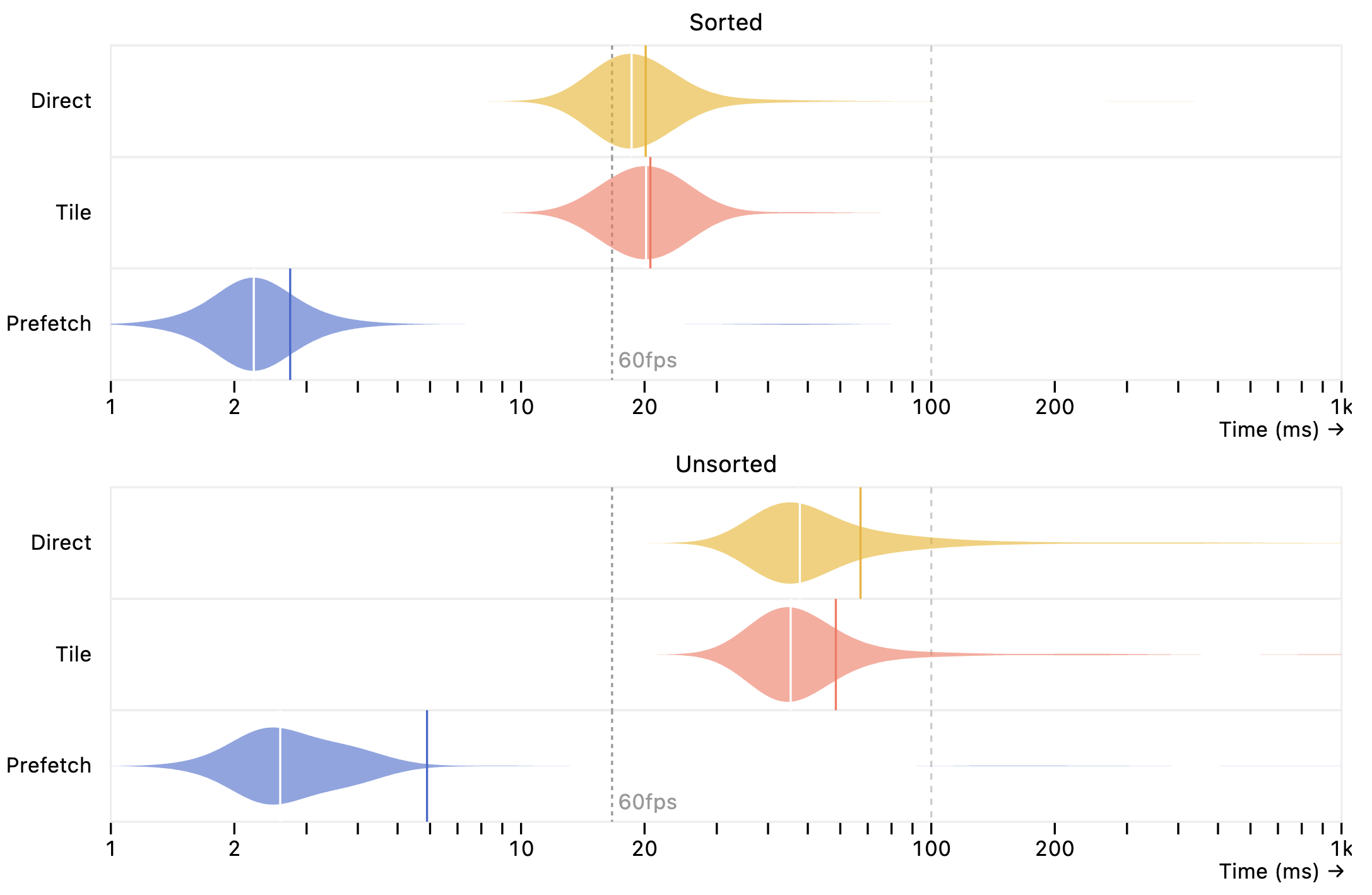}\vspace{-6pt}
\caption{Update latencies when panning the neuron recordings. Violin plots convey the latency update distribution over a log domain. White lines depict the median, colored lines depict the mean. Pre-sorting the data to align with the panning dimension improves the mean update time by a factor of 2-3. In either case, pre-fetching adjacent tiles provides nearly an order of magnitude improvement.}\vspace{-10pt}
\label{fig:tile-results}
\end{figure}

Figure~\ref{fig:tile-results} plots query latency (on a log scale), grouped by optimization conditions.
Violin plots show per-condition latency distributions, with white lines indicating the median time and colored lines indicating the mean (average) time.
In addition, grey lines indicate thresholds for 60 frames per second (dotted line) and 10 frames per second (dashed).
In all cases the distributions have skew---with mean latencies higher than the median latency---due to a tail of longer-running queries.

Without any prefetching, sorting provides clear improvements: median latencies reduce by more than half, while mean latencies reduce by almost a factor of three.
The worst-case time improves significantly, as the long tail of slower queries is greatly reduced.

Prefetching provides even larger improvements, with the median latency improving by almost a full order of magnitude.
In addition, sorting still makes important contributions for longer running queries.
In our setup, the \emph{skip} interactions do not benefit from prefetching, leading to higher latencies as the needed data must be queried on-demand; these are visible as thin blue regions towards the right side of the \emph{prefetch} condition plots.
With sorting, these standard queries take around 50ms (20fps), a reasonable interactive rate.
Without sorting, these queries instead require over 100ms (less than 10fps), running afoul of interface design guidelines \cite{psych-hci}.

\subsection{Future Work}
\label{sec:a-future-work}

The results indicate that both sorting and prefetching can provide substantial reductions in interactive latency.
In the future, we plan to create a ``publishing'' module that analyzes a Mosaic application and performs up-front optimizations including sorting and projection (e.g., removing unneeded columns) prior to publication.
While prefetching is already achievable, Mosaic clients currently must orchestrate their own prefetching requests.
We hope to augment the Mosaic coordinator with general methods for tile creation, tile stitching, and prefetching logic, alongside selection analysis, to enable more seamless and automatic prefetching support.

\section{Pre-Aggregated Materialized View Sizes}
\label{sec:pre-aggregated-materialized-view-sizes}

\begin{table*}[t]
\centering
\begin{tabular}{llrrlrr}
Template & Interactor & Interactive Resolution & Pixel Size & Client & Client Bins & Max. View Rows \\
\hline
Airlines & Time Slider & 240 & n/a & Confidence Intervals & 26 & 6,240 \\
\arrayrulecolor{lightgray}
\hline
Property & Date Interval & 440 & 1 & Regression Line & 1 & 440 \\
\hline
\multirow{6}{*}{Flights} & \multirow{2}{*}{Delay Interval} & \multirow{2}{*}{540} & \multirow{2}{*}{1} & Time Histogram & 24 & 12,960 \\
& & & & Distance Histogram & 30 & 16,200 \\\cline{2-7}
& \multirow{2}{*}{Time Interval} & \multirow{2}{*}{540} & \multirow{2}{*}{1} & Delay Histogram & 25 & 13,500 \\
& & & & Distance Histogram & 30 & 16,200 \\\cline{2-7}
& \multirow{2}{*}{Distance Interval} & \multirow{2}{*}{540} & \multirow{2}{*}{1} & Delay Histogram & 25 & 13,500 \\
& & & & Time Histogram & 24 & 12,960 \\
\hline
\multirow{12}{*}{Gaia} & \multirow{3}{*}{Map Interval 2D} & \multirow{3}{*}{$282 \times 178$ = 50,196} & \multirow{3}{*}{2} & Magnitude Histogram & 22 & 1,104,312 \\
& & & & Parallax Histogram & 25 & 1,254,900 \\
& & & & Color/Magnitude Plot & 50,196 & 2,519,638,416 \\\cline{2-7}
& \multirow{3}{*}{Magnitude Histogram} & \multirow{3}{*}{235} & \multirow{3}{*}{1} & Sky Map & 50,196 & 11,796,060 \\
& & & & Parallax Histogram & 25 & 5,875 \\
& & & & Color/Magnitude Plot & 24 & 5,640 \\\cline{2-7}
& \multirow{3}{*}{Parallax Histogram} & \multirow{3}{*}{235} & \multirow{3}{*}{1} & Sky Map & 50,196 & 11,796,060 \\
& & & & Magnitude Histogram & 22 & 5,170 \\
& & & & Color/Magnitude Plot & 50,196 & 11,796,060 \\\cline{2-7}
& \multirow{3}{*}{Plot Interval 2D} & \multirow{3}{*}{$178 \times 282$ = 50,196} & \multirow{3}{*}{2} & Sky Map & 50,196 & 2,519,638,416 \\
& & & & Magnitude Histogram & 22 & 1,104,312 \\
& & & & Parallax Histogram & 25 & 1,254,900 \\
\hline
\multirow{6}{*}{Taxis} & \multirow{2}{*}{Pick-Up Interval 2D} & \multirow{2}{*}{$168 \times 275$ = 46,200} & \multirow{2}{*}{2} & Time Histogram & 24 & 1,108,800 \\
& & & & Drop-Off Map & 46,200 & 2,134,440,000 \\\cline{2-7}
& \multirow{2}{*}{Drop-Off Interval 2D} & \multirow{2}{*}{$168 \times 275$ = 46,200} & \multirow{2}{*}{2} & Time Histogram & 24 & 1,108,800 \\
& & & & Pick-Up Map & 46,200 & 2,134,440,000 \\\cline{2-7}
& \multirow{2}{*}{Time Histogram} & \multirow{2}{*}{620} & \multirow{2}{*}{1} & Pick-Up Map & 46,200 & 28,644,000 \\
& & & & Drop-Off Map & 46,200 & 28,644,000
\end{tabular}\vspace{-6pt}
\caption{Interactive resolutions and client bin counts for each pre-aggregated materialized view in our benchmarks. The product of these values determines the maximum row count for a materialized view, corresponding to the memory requirements of a dense represention. The pixel size column corresponds to the \texttt{pixelSize} parameter of an interactor; for example, the actual screen pixel size of the Gaia sky map is $564 \times 356$, but the interval interactor (\texttt{pixelSize} = 2) and backing client raster grid both use a reduced resolution.}\vspace{-6pt}
\label{tbl:view-sizes}
\end{table*}

\begin{figure*}[t]
\centering
\includegraphics[width=\linewidth]{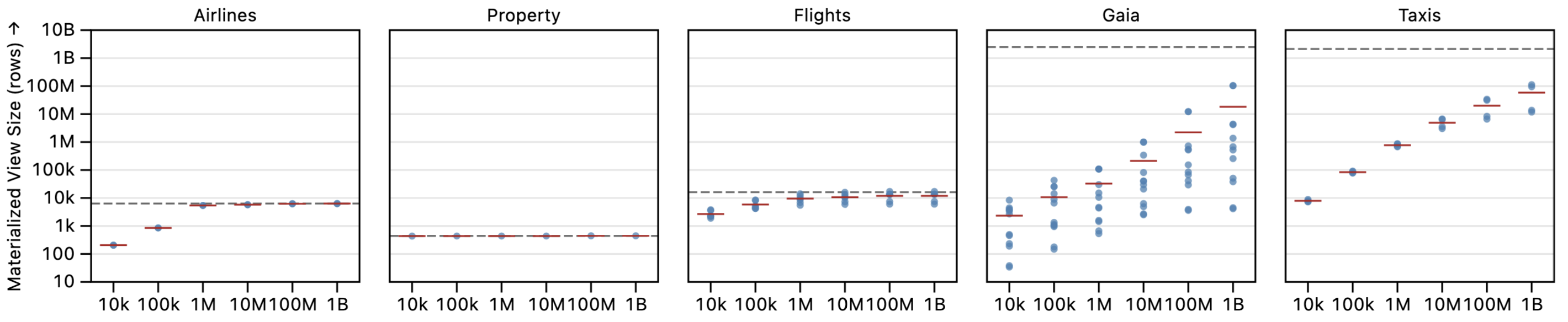}\vspace{-6pt}
\caption{Pre-aggregated materialized view sizes (row counts) across visualization and dataset size conditions. Points repesent single materialized views, red lines show averages, and dotted lines show the maximum possible size of a single materialized view. View sizes for histograms have lower maximum sizes and become dense as dataset size increases. Materialized views for high-resolution rasters are larger and sparser.}\vspace{-10pt}
\label{fig:sizes}
\end{figure*}

While the performance benchmarks in \S\ref{sec:eval} focus on query latency, here we further analyze the sizes of generated pre-aggregated materialized views.
Each pre-aggregated materialized view contains data for a single interactor/client pair.
The materialized view dimensions are determined by the \emph{interactive resolution} of the interactor (i.e., the number of interactive pixels for an \emph{interval} selection or the number of discrete values for a \emph{point} selection) and the number of discrete bins the client view uses to visualize the data.
The maximum view size is the product of these values; however, in practice a materialized view may be much smaller due to sparsity, where some of the bin combinations may not contain any data.

Table~\ref{tbl:view-sizes} summarizes the interactor and client bin counts for each of the visualization templates used in our benchmarks, including the maximum possible size of a single materialized view.
In the case of basic binned charts such as histograms, the maximum size is bounded around thousands of rows.
However, for higher-resolution plots such as raster visualizations, the maximum size increases dramatically, particularly when using an interval selection in one raster plot to filter a separate raster plot.
Optimization methods that use a dense representation \cite{doi:10.1111/cgf.12129, doi:10.1145/3290605.3300924} are not applicable in such cases due to insufficient memory to allocate such large dense arrays.

Figure~\ref{fig:sizes} plots the actual pre-aggregated materialized views for each visualization template and dataset size.
Each point represents a single materialized view, corresponding to one of the rows in Table~\ref{tbl:view-sizes}.
View sizes for the Airlines (slider + confidence intervals), Property (regression) and Flights (histograms) templates have lower maximum sizes and become quite dense as dataset size increases.
Meanwhile, materialized views for high-resolution raster displays (as in the Gaia and Taxis templates) are larger and sparser.
Though often an order of magnitude smaller than the input dataset size, cross-raster materialized views become too large for use in a browser via DuckDB-WASM with larger dataset sizes of 100M+ rows (as indicated earlier in Figure~\ref{fig:benchmark-results}).

One avenue for future work is to develop optimized schemes for managing pre-aggregated materialized views.
If view storage cost is a concern, a simple first step is to apply a basic LRU (least-recently used) caching policy, subject to a maximum storage constraint.
Additional optimization methods might estimate materialized view sizes for a given dataset and visualization specification, then use this information to dynamically adapt spatial and/or interactive resolution (interactor \texttt{pixelSize}) to bound the view size; trading off reduced resolution for storage cost and query speed improvements.

\end{document}